\documentclass[a4paper,11pt]{article}
\pdfoutput=1 
\usepackage{jcappub} 
\usepackage[T1]{fontenc} 

\usepackage{amssymb}
\usepackage{amsmath}
\usepackage{pifont}

\newcommand{\sg}{\delta \phi_{\rm r}}

\newcommand{\bm}[1]{\hbox{\boldmath{$#1$}}}
\newcommand{\sbm}[1]{\hbox{\boldmath{\scriptsize$#1$}}}

\newcommand{\revise}[1]{#1}

\title{\boldmath New scenario of QCD axion clump formation I: Linear analysis}

\author[a,b]{Naoya Kitajima,}
\author[c]{Kazuhiro Kogai,}
\author[c,d]{Yuko Urakawa}

\affiliation[a]{Frontier Research Institute for Interdisciplinary Sciences, Tohoku University,\\ Aramaki aza aoba, Aoba-ku, Sendai, 980-8578 Japan}
\affiliation[b]{Department of Physics, Tohoku University, \\Aramaki aza aoba, Aoba-ku, Sendai, 980-8578 Japan}
\affiliation[c]{Department of Physics and Astrophysics, Nagoya University, \\Chikusa, Nagoya 464-8602, Japan}
\affiliation[d]{Institute of Particle and Nuclear Studies, \\High Energy Accelerator Research Organization (KEK),\\ Oho, Tsukuba 305-0801, Japan}

\emailAdd{naoya.kitajima.c2@tohoku.ac.jp}
\emailAdd{kogai@nagoya-u.jp}
\emailAdd{yukour@post.kek.jp}

\abstract{
The QCD axion acquires the potential through the non-perturbative effect of the QCD matters around the QCD phase transition. During this period, the direct interaction between the axion and the QCD matters sets in. 
\revise{Focusing on the impact of this direct interaction, we propose two scenarios where the fluctuation of the axion can rapidly grow, potentially leading to the formation of axion miniclusters even if the Peccei-Quinn (PQ) symmetry was already broken during inflation. The first scenario assumes that the primordial curvature perturbation at the horizon scale during the QCD epoch was significantly enhanced and the second one assumes that the initial misalignment was tuned around the hilltop of the potential.}}


\begin{document}
\begin{flushright}
    KEK-TH-2367, 
    KEK-Cosmo-0281,
    TU1139
\end{flushright}
\maketitle
\flushbottom

\section{Introduction} \label{sec:intro}
Cosmological observations have revealed that our Universe is filled with unknown matter component called dark matter. The QCD axion is one of the most promising candidates for the dark matter. It was originally introduced in the Peccei-Quinn (PQ) mechanism as a solution of the strong CP problem \cite{Peccei:1977hh,Peccei:1977ur}. The QCD axion appears as a consequence of the PQ symmetry breaking \cite{Weinberg:1977ma,Wilczek:1977pj} and it acquires the mass through the QCD instanton effect at the QCD phase transition. After that, the QCD axion behaves as a cold dark matter component \cite{Preskill:1982cy,Abbott:1982af,Dine:1982ah}. For reviews, see e.g. \cite{Kim:2008hd,Kawasaki:2013ae,Marsh:2015xka,DiLuzio:2020wdo,Choi:2020rgn}.

It has been considered that the spatial distribution of the QCD axion below the cosmological scales, which can be precisely measured through the observations of the cosmic microwave background and the large scale structures, significantly differs, depending on whether the PQ symmetry was already broken during inflation or was broken after inflation. The former corresponds to the case when the PQ symmetry breaking scale, $f$, is comparable to or larger than the energy scale of inflation, characterized by the Gibbons-Hawking temperature, $T_{\rm GH} = H_I/2\pi$, with $H_I$ being the Hubble parameter during inflation, while the latter corresponds to the case when $f$ is smaller than $T_{\rm GH}$.

For $f<H_I/2\pi$,  which is called post-inflationary scenario, the spontaneous symmetry breaking occurs after inflation, resulting in the formation of topological defects \cite{Sikivie:1982qv}. Numerical simulations \cite{Fleury:2015aca, Vaquero:2018tib} have revealed that after the decay of topological defects, there remain a number of overdense axion clumps~\cite{Kolb:1993zz, Kolb:1993hw, Kolb:1994fi}. The axion overdense clumps undergo the non-linear gravitational collapse mainly after the matter-radiation equality, forming axiton~\cite{Kolb:1993hw} or axion miniclusters \cite{Hogan:1988mp}, which are gravitationally bound \cite{Zurek:2006sy}. More recently, in Ref.~\cite{Eggemeier:2019khm}, the formation and subsequent evolution of axion minihalos were studied based on $N$-body simulation to date (see also Refs.~\cite{Tinyakov:2015cgg, Kavanagh:2020gcy, Xiao:2021nkb}). A semi-analytic computation of the mass function for axion miniclusters can be found in Ref.~\cite{Enander:2017ogx}. For a review on axion miniclusters, see e.g., Ref.~\cite{Tkachev:2015usb}.

Observational consequences of axion miniclusters have been explored. Axion miniclusters can cause gravitational lensing effect which may be detectable through femtolensing and picolensing measurements~\cite{Kolb:1995bu} and microlensing measurements \cite{Fairbairn:2017dmf,Fairbairn:2017sil,Dai:2019lud} (see also Refs.~\cite{Fujikura:2021omw}). In Ref.~\cite{Tkachev:2014dpa}, it was argued that axion miniclusters may cause fast radio burst through their explosive decay. Meanwhile, in Refs.~\cite{Kolb:1993zz, Seidel:1993zk}, it was conjectured that axion stars, which are gravitationally bound and stable, may appear in the center of axion miniclusters through further contraction (see also Refs.~\cite{Braaten:2015eeu, Levkov:2016rkk, Visinelli:2017ooc,Chavanis:2017loo,Niemeyer:2019aqm,Eggemeier:2019jsu, Hertzberg:2020hsz}).

On the other hand, when the PQ symmetry is already broken during inflation, satisfying $f > H_I/2\pi$, the rapid expansion during inflation makes the axion field value almost homogeneous at least in our observable patch of the Universe, allowing only a tiny fluctuation of the initial misalignment. Because of that, it has been widely considered that a discovery of axion miniclusters can provide a smoking gun of the post-inflationary scenario. Nevertheless, in order to verify this claim, we need to exclude the possibility that axion miniclusters are formed for $f > H_I/2\pi$.

Lately, this possibility has been addressed more carefully for the QCD axion in Refs.~\cite{Arvanitaki:2009fg, Fukunaga:2020mvq, Sikivie:2021trt} and also for the axion like particles in Refs.~\cite{Hardy:2016mns, Blinov:2019jqc}. In Refs.~\cite{Arvanitaki:2009fg, Fukunaga:2020mvq}, it was shown that the self-interaction of the axion can significantly enhance the fluctuation of the axion around the QCD phase transition, at which the (misalignment) axion with a phenomenologically viable parameter commenced the oscillation. In this paper, we will point out that the direct interaction between the QCD axion and the QCD matters, which was ignored in Refs.~\cite{Arvanitaki:2009fg, Fukunaga:2020mvq}, drastically changes the dynamics of the axion in the QCD epoch. In Ref.~\cite{Sikivie:2021trt}, which has appeared almost at the same time as this paper, Sikivie and Xue also have considered the impact of the direct interaction on the fluctuation of the axion, arriving at a different conclusion from ours. We will explain the reason, focusing on a distinct difference between the fluctuation of the field value and the one of the energy density for the axion. Finally, we will point out new possible scenarios which may potentially lead to a formation of axion miniclusters even when the PQ symmetry was already broken during inflation.

This paper is organized as follows. In Sec.~\ref{sec:property}, we overview the basic properties of the QCD axion and the background evolution in the QCD epoch. In Sec.~\ref{Sec:Linearanalysis}, we summarize the basic equations to be solved at the linear order of perturbation. We also show the fluctuation of the axion follows a rather different evolution at superhorizon scales, depending on the gauge choice. In Sec.~\ref{Sec:Linear}, we first summarize the possible mechanisms which can prominently enhance the inhomogeneity of the axion. Subsequently, solving the equations derived in Sec.~\ref{Sec:Linearanalysis}, we compute the linear evolution of the fluctuation of the axion. We also point out two possible scenarios which may lead to a formation of axion miniclusters. Finally, in Sec.~\ref{sec:conc}, we summarize our result and future issues. In this paper, we take the present dark matter density $\Omega_{\rm DM} = 0.23$ and the Hubble constant $H_0 = 67.8\,{\rm km/s/Mpc}$. We use the reduced Planck mass $M_{\rm pl} = 1/\sqrt{8\pi G}$.

\section{Background evolution of QCD axion}
\label{sec:property}
In this section, we overview the background evolution of the axion in the QCD epoch.

\subsection{Axion potential}
At the QCD phase transition, the QCD axion acquires the potential through the QCD instanton effect. The height of the potential is determined by the QCD scale. Here we adopt the results of the lattice QCD simulation in Ref.~\cite{Borsanyi:2016ksw}, where the QCD topological susceptibility $\chi$ is approximately given by
\begin{equation}
    \chi(T)=\frac{\chi_0}{1+\tilde{T}^b}, \qquad \quad \tilde{T} \equiv T/T_{\rm c} \label{eq:chiQCD}
\end{equation}
with $\chi_0=(7.6\times10^{-2}\ {\rm GeV})^4$, $b=8.2$, and $T_{\rm c} = 0.15$\,GeV. Here, $T$ should be understood as the temperature of the relativistic species, which dominate the early Universe. For our later use, let us introduce the dimensionless axion field and the dimensionless conformal time $\eta$ as 
\begin{align}
   \tilde{\phi} \equiv  \frac{\phi}{f}\,, \qquad  \eta \equiv  \frac{T_{\rm c}}{T} =\frac{a}{a_{\rm c}}\,,
\end{align}
where $\eta$ is related to $t$ as $H_{\rm c} dt = \eta d \eta$. When the dilute instanton gas approximation holds, the potential of the QCD axion, $\phi$, is given by 
\begin{equation}
    V(\phi)=\chi(T)(1-\cos(\phi/f)) \equiv \chi(T) \tilde{V} (\tilde{\phi}).
    \label{eq:potcos}
\end{equation}
The potential form slightly differs from the one computed from the low energy effective Lagrangian in Ref.~\cite{DiVecchia:1980yfw}. A detailed study about the potential form of the QCD axion and the coupling with the standard model sector can be found, e.g., in Refs.~\cite{diCortona:2015ldu, Fox:2004kb, Berkowitz:2015aua, Borsanyi:2015cka}.

\subsection{Background evolution}
The background equation of motion for the axion is given by
\begin{align}
    \ddot{\phi} + 3 H \dot{\phi} + V_{\phi} = 0\,.  \label{Eq:BGKG}
\end{align}
Here and hereafter, the lower indices $X$ put on $V$ denote the partial differentiation with respect to $X$.
For $T > T_{\rm c}$, as the temperature decreases, the mass of axion, given by
\begin{equation}
    m(T)= \sqrt{\chi(T)}/f\,,
\end{equation}
increases as $m(T)\propto T^{-b/2}$. Meanwhile, for $T < T_{\rm c}$, $m(T)$ approaches to a constant value, given by
\begin{equation}\label{eq:axionmass}
m_0 \simeq 6\,\mu{\rm eV}\left(\frac{10^{12}\,{\rm GeV}}{f} \right).
\end{equation} 
In this paper, we assume that the QCD matter, quarks and gluons, are in thermal equilibrium, having the same temperature $T$ as other relativistic components. Under this assumption, we indistinguishably use the words QCD matter and radiation in this paper.  

\begin{figure}
\centering
\includegraphics[width=0.48\linewidth]{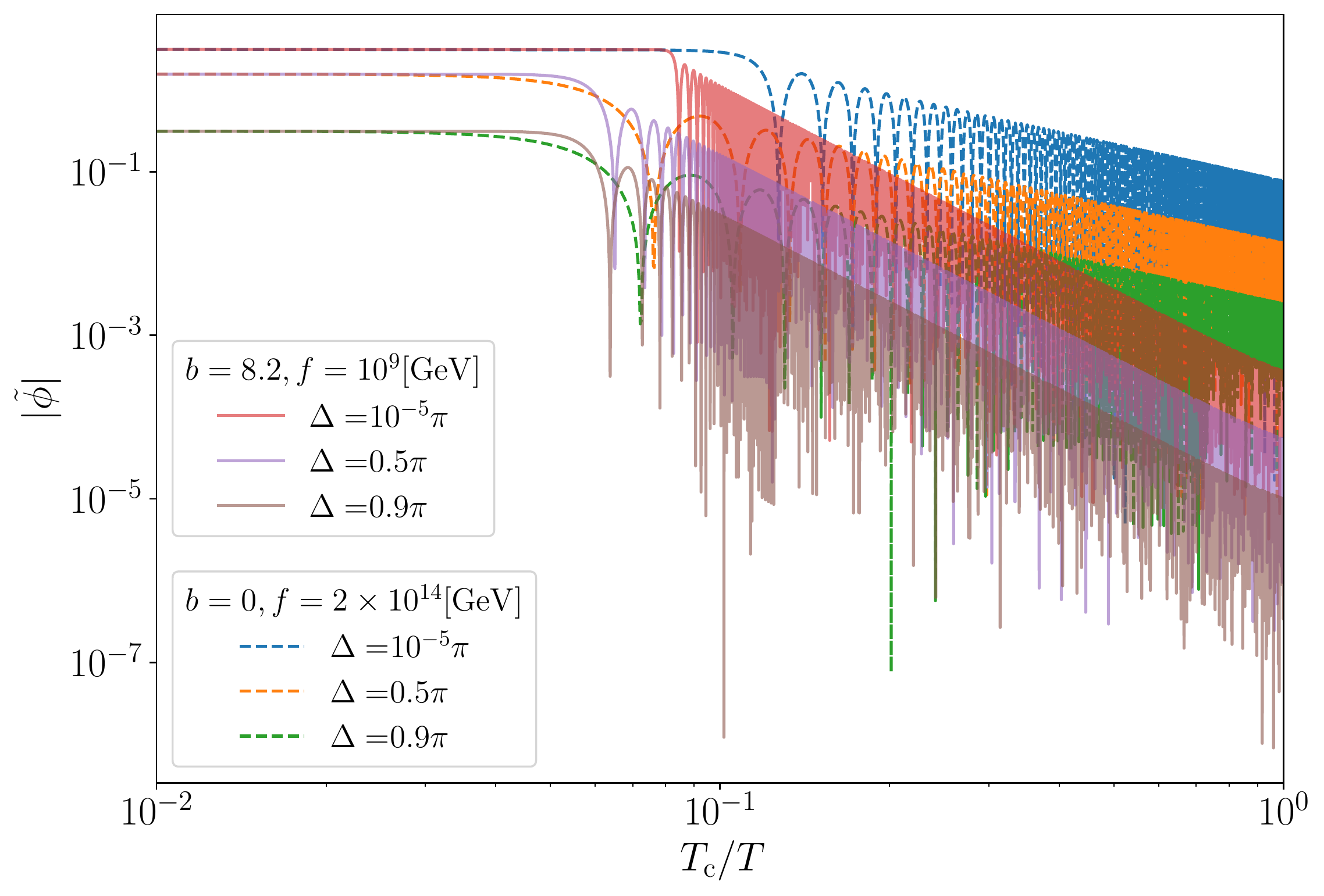}
\includegraphics[width=0.48\linewidth]{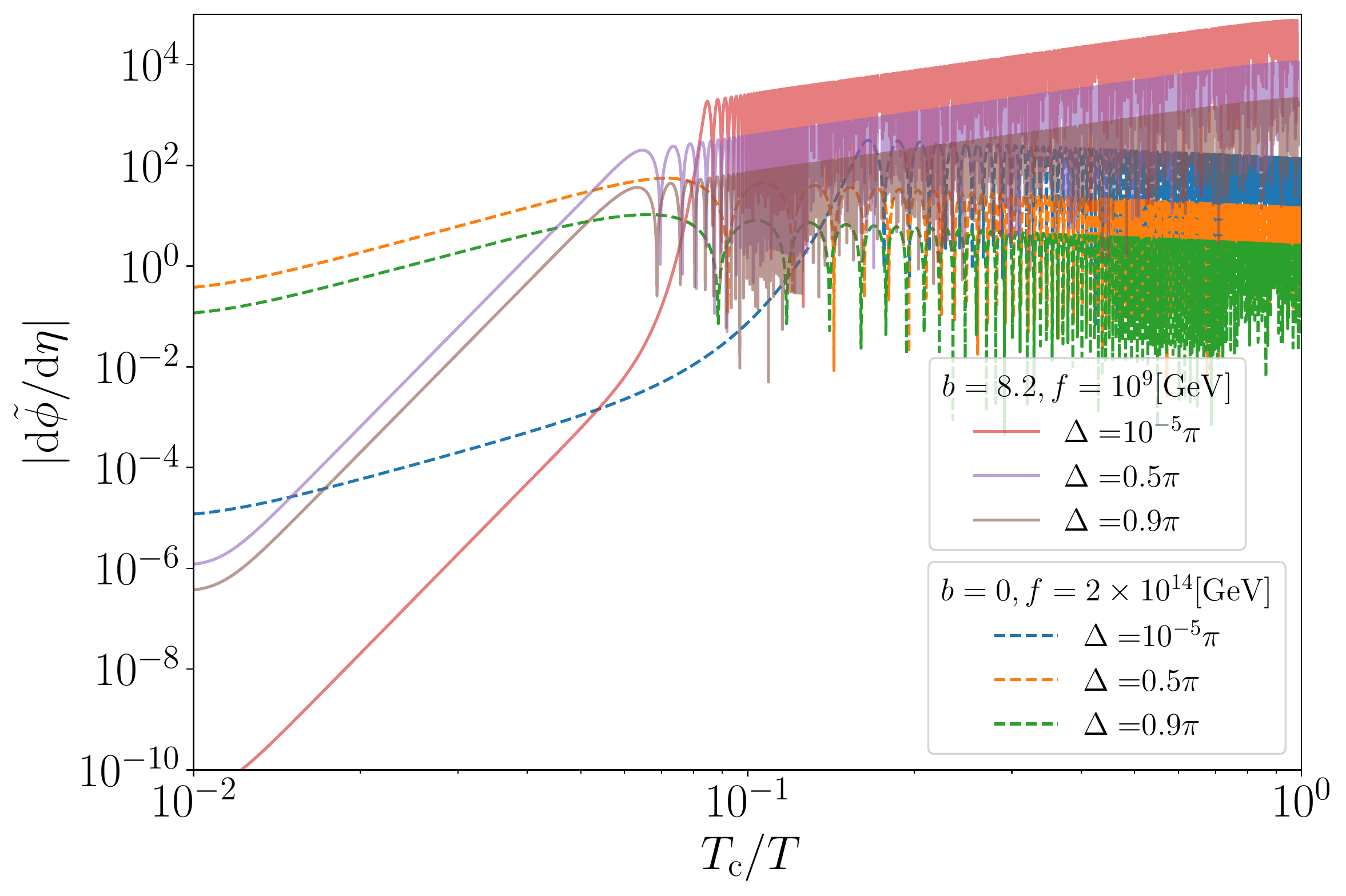}
\caption{These plots compare the background evolution for $b=8.2$ with the temperature dependent mass and for $b=0$ with the constant mass for various initial conditions, characterized by $\Delta \equiv \pi - \phi_{\rm i}/f$. The initial velocity is determined by imposing the slow-roll condition. The right panel shows the evolution of $d \tilde{\phi}/d \eta$. The PQ scale is set to $f=10^{9}\,{\rm GeV}$ for $b=8.2$ and $f = 2\times10^{14}\,{\rm GeV}$ for $b=0$. The initial velocity is determined by imposing the slow-roll condition. \label{Fg:background}}
\end{figure}
\begin{figure}
    \centering
    \includegraphics[width=0.8\linewidth]{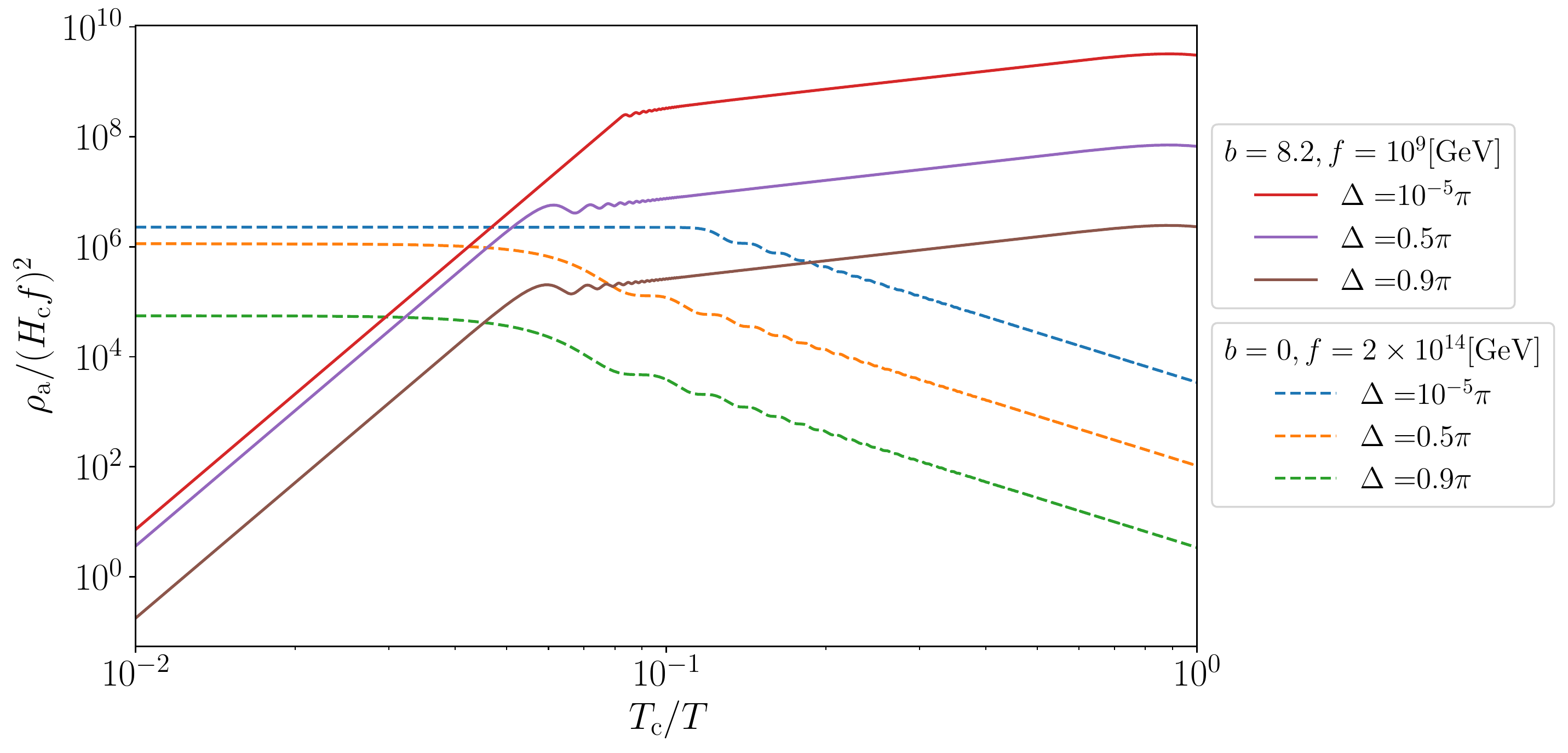}
    \caption{ 
    The axion background density. The solid lines are $b=8.2$ and the dashed lines are $b=0$. From the top to bottom, $\Delta = 10^{-5}\pi, 0.5\pi, 0.9\pi$.
    We chose the decay constant $f$ for $b=0$ so that the axion commences the oscillation almost at the same time as the case with $b=8.2$.}
    \label{fig:rho}
\end{figure}

Figure \ref{Fg:background} shows the background evolution of $\tilde{\phi}$ and $d \tilde{\phi}/d \eta \, (\equiv \tilde{\phi}')$ for $b= 0$ and $b= 8.2$. We characterize the initial misalignment $\tilde{\phi}_{\rm i}$, using 
\begin{align}
    \Delta \equiv \pi - \phi_{\rm i}/f = \pi - \tilde{\phi}_{\rm i}\,,
\end{align}
with $0 < \Delta < \pi$.
For $b=8.2$, the slow-roll condition is more abruptly violated around $H \sim m$. While the velocity $d \tilde{\phi}/d \eta$ decreases as $a^{-1/2} \propto \eta^{- 1/2}$ for $b=0$ due to the Hubble friction during the oscillation, it increases for $b= 8.2$, since the transferred energy from the QCD matter overcomes the decrease due to the Hubble friction. \revise{Figure \ref{fig:rho} shows the evolution of the background axion density in the same setup as figure \ref{Fg:background}.}

According to the numerical analysis, during the harmonic oscillation, the overall amplitude of $d\tilde{\phi}/d \eta$ and $\tilde{V}_{\tilde{\phi}}$ evolve as 
\begin{align}
    \frac{d \tilde{\phi}}{d \eta} \propto \eta^{1.5}\,, \qquad \tilde{V}_{\tilde{\phi}} \propto \eta^{- 3.6}
\end{align}
for $T \geq T_{\rm c}$. To analytically derive this overall evolution, let us solve
\begin{align}
    \dot{\rho}_{\rm a} + 3 H (\rho_{\rm a} + p_{\rm a}) = V_\rho \dot{\rho} = - \frac{d \ln \chi}{d \ln T}\, V H \,, \label{Eq:ecsv_bgav}
\end{align}
which can be derived by using Eq.~(\ref{Eq:BGKG})\footnote{The background energy density and the pressure for the axion read
\begin{align}
    \rho_{\rm a} = \frac{1}{2} \dot{\phi}^2 + V(\phi,\, \rho)\,, \qquad  p_{\rm a} = \frac{1}{2} \dot{\phi}^2 - V(\phi,\, \rho) \label{Exp:EMbg}\,.
\end{align}
Using the background Klein-Gordon equation (\ref{Eq:BGKG}), we obtain Eq.~(\ref{Eq:ecsv_bgav}).}. On the second equality, we have used the energy conservation for the radiation with $p = \rho/3$, where $p$ and $\rho$ denote the pressure and the energy density of the radiation. For simplicity, let us further assume that $d \ln \chi/d \ln T$ remains almost constant, $-b$, for $T> T_{\rm c}$ and for $T< T_{\rm c}$, respectively. 

During the harmonic oscillation, using $p_{\rm a} \simeq 0$ and $V \simeq \rho_{\rm a}/2$, Eq.~(\ref{Eq:ecsv_bgav}) can be integrated to give
\begin{align}
    \rho_{\rm a}  \propto a^{-3(1- \frac{b}{6})} \propto \eta^{-3(1- \frac{b}{6})}\,,
\end{align}
which in turn implies
\begin{align}
    \frac{d \tilde{\phi}}{d \eta} \propto \eta^{- \frac{1}{2} (1 - \frac{b}{2})}\,, \qquad \tilde{V}_{\tilde{\phi}} \sim \tilde{\phi} \propto \eta^{- \frac{1}{4}(6+b)}\,,  \label{Exp:backgroundscaling}
\end{align}
reproducing the numerical result approximately. For $\Delta \ll 1$, due to the anharmonic correction, the evolution of the axion differs from the one discussed here right after the onset of the oscillation.  

\subsection{Relic abundance}
When the initial misalignment is not particularly large, the axion commences the oscillation at around $T_{\ast}$ with $3H(T_\ast) = m(T_\ast) \equiv m_\ast$, where $H(T)$ is the Hubble parameter. For $T_\ast > T_{\rm c}$, $T_\ast$ is given by
\begin{equation}
T_\ast \simeq 1.0\,{\rm GeV} \left(\frac{10^{12}\,{\rm GeV}}{f}\right)^{0.16}.
\end{equation}
Then, one can obtain the final QCD axion abundance as (see also Refs.~\cite{Turner:1985si, Fox:2004kb,Visinelli:2009zm}) 
\begin{equation}
\Omega_a h^2 = 0.10 \kappa \left(\frac{f}{10^{12}\,{\rm GeV}} \right)^{1.16}\, \tilde\phi_{\rm i}^2\, F(\tilde\phi_{\rm i}).
\end{equation}
Here, $\kappa$ is ${\cal O}(1)$ numerical factor which characterizes nonadiabaticity and $F(\tilde\phi_{\rm i})$ is anharmonic correction factor which is a function of the initial value of $\tilde\phi = \phi/f$. The anharmonic correction $F(\tilde\phi_{\rm i})$ was computed, e.g., in Refs.~\cite{Turner:1985si, Lyth:1991ub, Bae:2008ue}. Note that the QCD axion with the decay constant being $f \sim 10^{12}$\,GeV and the initial value being $\tilde\phi_{\rm i} \sim 1$ saturates the DM relic abundance in the Universe, corresponding to the upper bound of the so called "classical axion window"~\cite{ABBOTT1983133}, given by  
\begin{equation}
    10^{8}{\rm GeV} \lesssim f \lesssim 10^{12}{\rm GeV},  \label{Eq:rangef}
\end{equation}
where the lower bound comes from neutrino burst duration of SN1987A~\cite{Mayle:1987as}\footnote{Since the QCD axion is introduced to solve the strong CP problem, the QCD axion should generically interact with the neutron electronic dipole moment. 
Then, the observation of the SN1987A neutrino burst gives the lower bound of the decay constant $f\gtrsim9\times10^5 {\rm GeV}$ for generic QCD axion models\,\cite{Graham:2013gfa,DiLuzio:2020wdo}.}. The detailed values of the upper bound and the lower bound depend on the models of the QCD axion and also on the cosmological scenario.

Meanwhile, when the axion was initially located around the potential maximum, taking a large initial misalignment, as considered in Refs.~\cite{Fukunaga:2019unq, Arvanitaki:2019rax}, the abundance constraint becomes much tighter than the one in Eq.~(\ref{Eq:rangef}). In what follows, we calculate the abundance constraint on $f$, using the adiabatic invariant
\begin{align}
    I(T) \equiv \frac{\rho_{\rm a}(T)/m(T)}{s(T)}\,, 
\end{align}
which remains constant during an adiabatic evolution~\cite{Bae:2008ue}. Our numerical computation shows that $I(T)$ becomes almost constant after several cycles of oscillation as well as for $\tilde{\phi}_{\rm i} \simeq \pi$. Using $I(T)$, let us define
\begin{align}
    \hat{I}(T) \equiv \frac{s_0 m_0 h^2}{\rho_{\rm cr}} I(T) = \frac{s_0 m_0}{3 M_{\rm pl}^2} \left( \frac{h}{H_0} \right)^2 I(T)\,. 
\end{align}
Once $\hat{I}$ approaches to a constant value, it directly gives the present axion abundance as
\begin{align}
   \Omega_{\rm a} h^2 = \frac{1}{\gamma}  \hat{I}(T_{**})\,,
\end{align}
where $T_{**}$ denotes a reference temperature after $I(T)$ is settled down to a constant value and $\gamma$ characterizes a possible change of $I(T)$ after $T_{**}$ as
\begin{align}
    I (T_{**}) = \gamma I(T_0)\,. 
\end{align}

\begin{figure}
    \centering
    \includegraphics[width=0.49\linewidth]{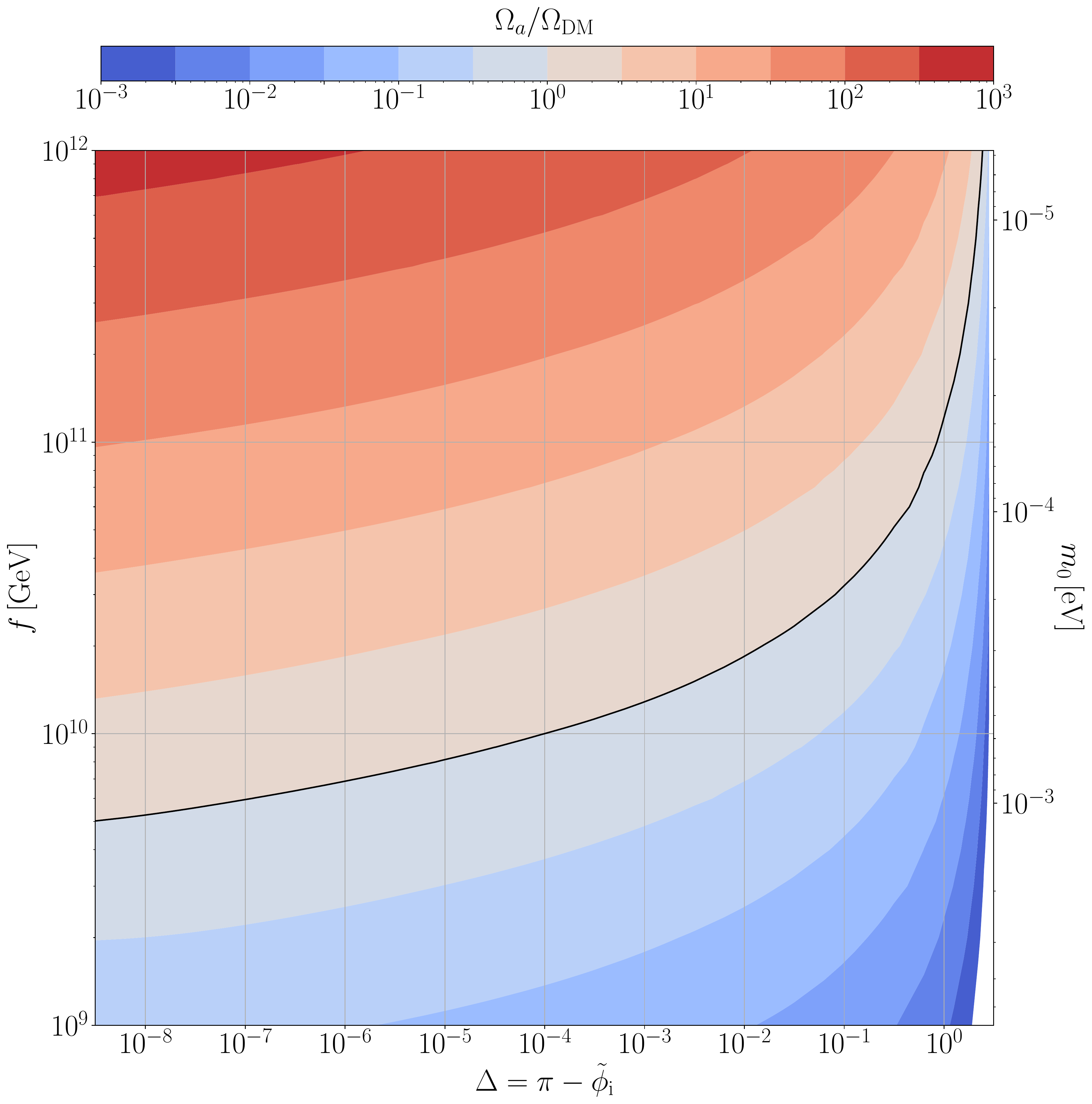}
    \includegraphics[width=0.49\linewidth]{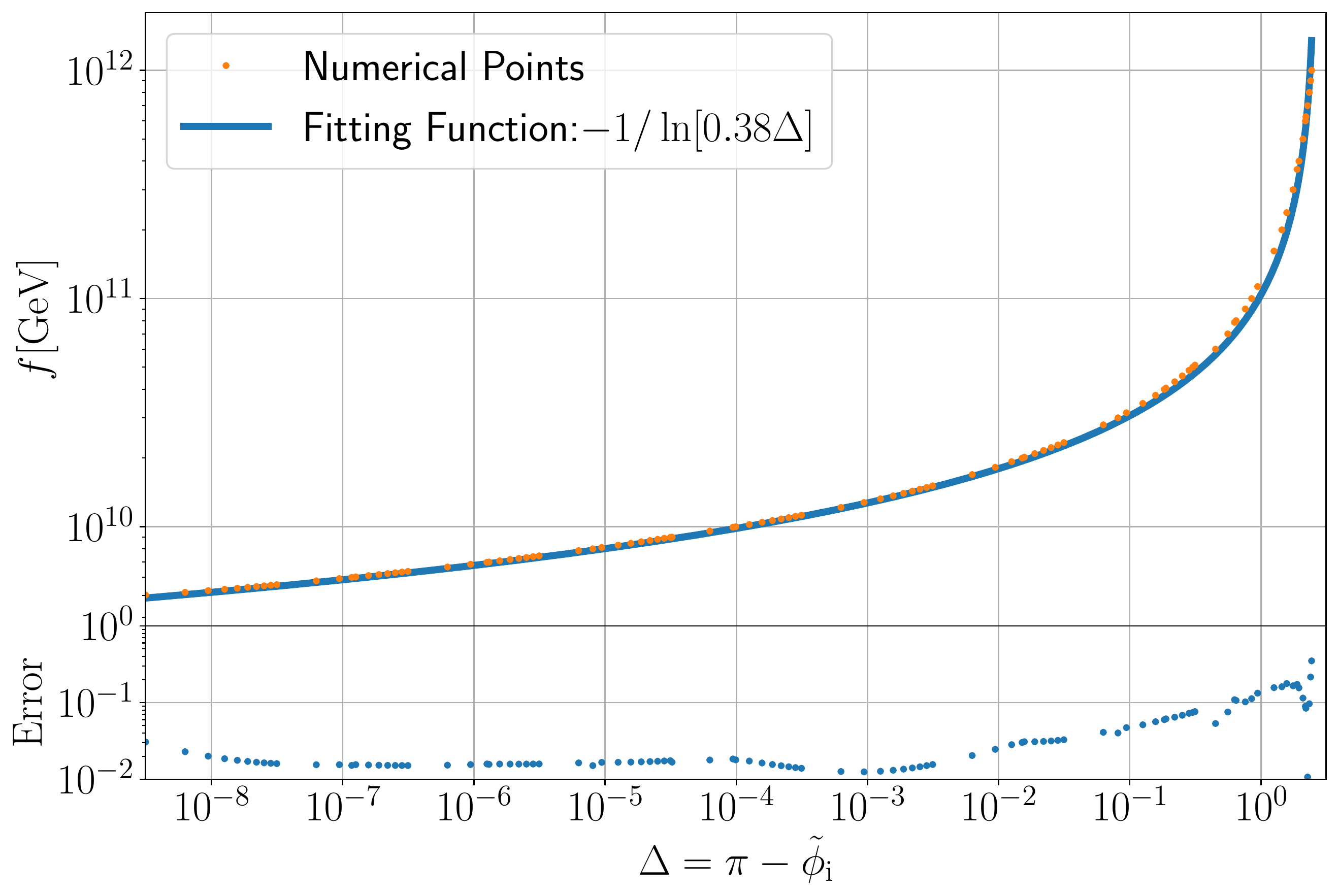}
    \caption{\label{fig:abundance}The left panel shows the ratio of the axion abundance to the dark matter. The black curve corresponds to $\Omega_a=\Omega_{\rm DM} = 0.23$. The horizontal axis denotes the initial deviation from the potential maximum, $\Delta$. The right panel shows the corresponding value of $f$ for a given $\Delta$, when the QCD axion saturates the dark mater abundance, corresponding to the black curve in the left panel.}
\end{figure}

The left panel of Fig.~\ref{fig:abundance} shows the axion abundance for a given PQ scale $f$ (or a axion mass $m$) and a given initial misalignment $\tilde{\phi}_{\rm i}$. The black line shows the contour with $\Omega_a h^2 = 0.11$, corresponding to the case when the QCD axion saturates the total dark matter abundance. The right panel shows the corresponding value of $f$ for each value of $\Delta$ for $\Omega_a h^2 = 0.11$. The orange dots are the numerical values and the blue curve is the fitting function, given by 
\begin{align}\label{eq:f-delta}
    \left(\frac{f(\Delta)}{10^{11}[{\rm GeV}]}\right) = -\frac{1}{\left[\log\left(0.38\Delta\right)\right]}\,,
\end{align}
which approximates the numerical result with a few \% error for $\Delta \ll 1$. This extends the analysis in Ref.~\cite{Bae:2008ue} to a smaller value of $\Delta$. An analytical derivation of the fitting function $f(\Delta)$ was intended in Ref.~\cite{Lyth:1991ub} based on a somewhat rough argument.

\section{Linear evolution of axion inhomogeneity}  \label{Sec:Linearanalysis}
In this section, we summarize the perturbed Klein-Gordon equation for the QCD axion in two different gauges: uniform radiation density slicing and Newtonian gauge. The former/latter is convenient to solve the superhorizon/subhorizon evolution, respectively. 

\subsection{Perturbed equations before gauge fixing}
Using the metric perturbations $A$, $B$, $D$, and $E$, we express the line element as
\begin{align}
    ds^2 = - (1 + 2 A Y) dt^2 - 2 a B Y_i dt dx^i + a^2 \left( (1+ 2 D Y) \delta_{ij} + 2 E Y_{ij} \right) dx^i dx^j \,,
\end{align}
where $a$ is the scale factor. Following Refs.~\cite{Bardeen:1980kt, Kodama:1985bj}, we expand the linear perturbations with a complete set of the scalar harmonic function $Y$ defined on the spatial slicing. In this paper, we assume the spatially flat background. Using these metric perturbations, we introduce the curvature perturbation
\begin{align}
   {\cal R} \equiv D + E/3 \,,
\end{align}
which gives the spatial Ricci scalar on a time constant surface, ${^s\!R}$, as ${^s\!R} = 4 (k/a)^2 {\cal R} Y$, and the shear for the normal vector to the constant time slicing, $\sigma_{\rm g}$ as
\begin{align}
  \sigma_{\rm g} \equiv \frac{a}{k} \dot{E} - B\,.   
\end{align}
Both of ${\cal R}$ and $\sigma_{\rm g}$ are not gauge invariant variables, since they transform as 
\begin{align}
    {\cal R} \to {\cal R} - H \delta t\,, \qquad \sigma_{\rm g} \to \sigma_{\rm g} - \frac{k}{a} \delta t\,,
\end{align}
when we change the time slicing as $t \to t+ \delta t$.

The perturbed equation at linear order is given by
\begin{align}
   & \delta  \ddot{\phi} + 3 H \delta  \dot{\phi} + V_{\phi \phi} \delta \phi + \left( \frac{k}{a} \right)^2 \delta \phi   \cr 
   & \qquad \qquad \qquad + V_{\phi \rho} \delta \rho  - \dot{\phi} \dot{A} + 2 V_{\phi} A + \dot{\phi} \left(3 \dot{\cal R}  - \frac{k}{a} \sigma_{\rm g} \right) = 0  \,,\label{Eq:KGp}
\end{align}
where $\delta \rho$ denotes the density perturbation of radiation. Since the potential $V$ also depends on the radiation temperature $T$ or the energy density of the radiation, we should include the first term in the second line. This term, which will turn out to be crucial both in the superhorizon and subhorizon scales, was not taken into account in Ref.~\cite{Arvanitaki:2019rax}. Using the Hamiltonian constraint, the density perturbation of radiation can be expressed as 
\begin{align}
    3 \left (A - \frac{\dot{\cal R}}{H} \right) + \frac{k}{aH} \sigma_{\rm g} - \left( \frac{k}{aH} \right)^2 {\cal R} = - \frac{4 \pi G}{H^2} \delta \rho_{\rm tot} \,, \label{Hconst}
\end{align}
where $\delta \rho_{\rm tot}$ denotes the fluctuation of the total energy density.

\subsection{Superhorizon evolution in uniform density slicing}  \label{SSec:superhorizon}
Let us start with discussing the superhorizon evolution, $k/(aH) \ll 1$. In this subsection, we show that when we choose an appropriate basis, the adiabatic perturbation does not affect the axion perturbation, which corresponds to the entropy perturbation, in superhorizon scales. This property was shown for a more general multi scalar field system in Ref.~\cite{Gordon:2000hv}.

For this purpose, let us introduce a gauge invariant perturbation for the axion perturbation as
\begin{align}
    \sg \equiv \delta \phi - \frac{\dot{\phi}}{\dot{\rho}} \delta \rho\,,  \label{Def:sg}
\end{align}
which corresponds to the axion fluctuation in uniform radiation density slicing with $\delta \rho=0$. The orthogonality between $\sg$ and the adiabatic perturbation becomes manifest by taking $\delta \rho =0$ slicing. In the absence of the anisotropic pressure, the shear $k \sigma_{\rm g}$ decays as $1/a^2$ in superhorizon scales. Therefore, in this slicing, the Hamiltonian constraint and the Klein-Gordon equation at linear perturbation read
\begin{align}
  &  A_{\rm r} - \frac{\dot{\zeta}}{H} + \frac{1}{2} \frac{\delta \rho_{\rm a, r}}{\rho_{\rm tot}} \simeq 0\,,  \label{Eq:HeqUR}\\
  &  \delta  \ddot{\phi}_{\rm r} + 3 H \delta  \dot{\phi}_{\rm r} + V_{\phi \phi} \sg  - \dot{\phi} \dot{A}_{\rm r} + 2 V_{\phi} A_{\rm r} + 3 \dot{\phi}  \dot{\zeta}  \simeq 0\,,  \label{Eq:KGeqUR}
\end{align}
where $\delta \rho_{\rm a, r}$ denotes the perturbation of the axion energy density, given by 
\begin{align}
\delta \rho_{{\rm a,r}} = \dot{\phi} \delta \dot{\phi}_{\rm r} + V_{\phi} \sg  - A_{\rm r} \dot{\phi}^2  \label{Exp:drhophi_UD}
\end{align}
and $\zeta$ denotes the curvature perturbation in this slicing, given by
\begin{align}
    \zeta \equiv {\cal R} - \frac{H}{\dot{\rho}} \delta \rho\,.
\end{align}
In Eqs.~(\ref{Eq:HeqUR}) and (\ref{Eq:KGeqUR}), we use $\simeq$, since the terms suppressed by $(k/aH)^2$ and the decaying contributions in an expanding Universe are ignored. The lapse perturbation $A_{\rm r}$ is related to the one in another slicing as
\begin{align}
    A_{\rm r} =  A - \partial_t \left( \frac{\delta \rho}{\dot{\rho}} \right) \,. \label{Exp:Ar}
\end{align}

When the potential of the axion depends on the temperature or the energy density of the radiation, the axion and the radiation have the corresponding direct interaction. Then, the energy-momentum tensors for the axion and the radiation are not independently conserved. Writing down the energy conservation equation for the radiation, which includes the energy flow from the QCD matters (radiation) to the axion, we obtain
\begin{align}
    \frac{\dot{\zeta}}{H} \simeq - \frac{\dot{\rho}}{4 H \rho} V_{\phi \rho} \sg \,, \label{Eq:EC}
\end{align}
where the term in the right hand side describes the energy flow, which vanishes once the potential of the axion ceases to depend on the temperature. Using Eqs.~(\ref{Eq:HeqUR}) and (\ref{Eq:EC}), we can eliminate $A_{\rm r}$ and $\zeta$ in Eq.~(\ref{Eq:KGeqUR}).

In the QCD epoch, the energy fraction of the axion is still much smaller than the one of the radiation, satisfying
$$
\frac{\rho_{\rm a}}{\rho} \bigg|_{T=T_{\rm c}} \leq \frac{\rho_{\rm m}}{\rho} \bigg|_{T=T_{\rm c}} \simeq \frac{T_{\rm eq}}{T_{\rm c}} \simeq 10^{-8}\,,
$$
where the inequality is saturated when the axion is the dominant component of the dark matter. When we further ignore the terms suppressed by $\rho_{\rm a}/\rho \simeq \rho_{\rm a}/\rho_{\rm tot}$, we obtain  
\begin{align}
      \delta  \ddot{\phi}_{\rm r} + 3 H \delta  \dot{\phi}_{\rm r} + V_{\phi \phi} \sg \simeq 0\,,  \label{Eq:deltaphir}
\end{align}
which coincides with the equation of motion for the axion in the absence of the metric perturbation in the superhorizon limit (see Fig.~\ref{fig:dphiSP} in Appendix~\ref{App:diff-gauges}).

\subsection{Newtonian gauge}
Next, we consider the Newtonian gauge, where the line element is given by
\begin{align}
    ds^2 = - (1- 2\Phi) dt^2 + a^2(t) (1 + 2 \Phi) d \bm{x}^2 \,. 
\end{align}
Here, the lapse perturbation is given by $-\Phi$, since the anisotropic pressure vanishes. Ignoring the non-linear contributions of the metric perturbations, the Klein-Gordon equation is given as
\begin{align}
    \partial_t^2 \phi + 3 H \partial_t \phi - (1 - 4\Phi) \frac{\partial^2_{{\sbm x}}}{a^2} \phi + 4 \partial_t \Phi \partial_t \phi + (1 - 2 \Phi) V_\phi (\phi,\, \rho) =0\,,   \label{eq:KG}
\end{align}
where the axion is kept non-perturbative. Perturbing also about the field fluctuation of the axion, we obtain 
\begin{align}
    \partial_t^2 \delta \phi + 3 H \partial_t \delta \phi + \left( \frac{k}{a} \right)^2 \delta \phi + 4 \dot{\phi} \partial_t \Phi +  V_{\phi\phi} \delta \phi - 2 V_{\phi} \Phi + V_{\phi\rho} \delta \rho=0\,,   \label{eq:KGLG}
\end{align}
which can be obtained also by choosing the Newtonian gauge in Eq.~(\ref{Eq:KGp}). Here and hereafter, for notational brevity, we simply express the background axion field and the axion perturbation in Newtonian gauge as $\phi$ and $\delta \phi$, distinguishing the later from $\sg$. Using the Hamiltonian constraint (\ref{Hconst}), we can express the perturbation of the radiation density, $\delta \rho$, as
\begin{align}
    3 H^2 \left (\Phi + \frac{\dot{\Phi}}{H} \right)  + \left( \frac{k}{a} \right)^2 \Phi =  4 \pi G \delta \rho_{\rm tot} \simeq   4 \pi G \delta \rho\,, \label{Eq:deltarho}
\end{align}
where on the second equality, we have ignored the perturbation of the axion energy density, which is suppressed by $\rho_{\rm a}/\rho$\footnote{This approximation should be more carefully examined, as it approaches to the matter-radiation equality. As is known, the density perturbation of dark matter dominates the one of radiation even before the equality \cite{Meszaros, Hu:1995en} for the slow solution~\cite{Weinberg:2002kg}.}.

Using the dimensionless conformal time $\eta$, we obtain the field equation for the axion perturbation in Newtonian gauge, $\delta \phi$, at linear perturbation as
\begin{align}
    &\partial_\eta^2 \delta \tilde{\phi} + \frac{2}{\eta} \partial_\eta \delta \tilde{\phi}+ \tilde{k}^2 \delta \tilde{\phi}  + \eta^2 \frac{\chi}{(H_{\rm c} f)^2} \tilde{V}_{\tilde{\phi} \tilde{\phi}} \delta \tilde{\phi}  = S_1 + S_2 + S_3  \,, \label{Eq:KGXRD}
\end{align}
where the source terms are given by
\begin{align}
    & S_1 \equiv - 4 \partial_\eta \Phi \partial_\eta \tilde{\phi}\,, \qquad S_2 \equiv  2 \Phi \eta^2  \tilde{V}_{\tilde{\phi}} \frac{\chi}{(H_{\rm c} f)^2} \,, \cr
    & S_3 \equiv \frac{1}{4} \tilde{V}_{\tilde{\phi}} \eta^3 \frac{\delta \rho}{\rho}  \frac{d}{d \eta} \frac{\chi}{(H_{\rm c} f)^2}  = \frac{1}{2}  \tilde{V}_{\tilde{\phi}}  \left( \Phi +  \eta \partial_\eta \Phi + \frac{\tilde{k}^2}{3}  \eta^2 \Phi  \right) \eta^3\frac{d}{d \eta} \frac{\chi}{(H_{\rm c} f)^2}\,, \label{Def:sources}
\end{align}
and the normalized wavenumber $\tilde{k}$ is defined as
$$
\tilde{k} \equiv k/(a_{\rm c} H_{\rm c})
$$
with $a_{\rm c}$ and $H_{\rm c}$ being the scale factor and the Hubble parameter at $T=T_{\rm c}$. Here, we have eliminated $\delta \rho$, using Eq.~(\ref{Eq:deltarho}). The source term $S_3$ appears as a consequence of the direct interaction between the QCD axion and QCD matters (radiation), which is encoded as the temperature dependence of the axion mass. Because of the non-vanishing direct interaction, the energy momentum tensors for the axion and radiation are not separately conserved for $T \geq T_{\rm c}$. Once the axion mass becomes independent of the temperature for $T < T_{\rm c}$, since the direct interaction vanishes, $S_3$ also vanishes. Meanwhile, the other source terms $S_1$ and $S_2$ appear due to the indirect interaction through gravity. Therefore, these terms do not vanish as well as after the axion mass becomes independent of the temperature.

When the total matter of the Universe is barotropic, satisfying the adiabatic condition, and the anisotropic pressure vanishes, the equation of motion for $\Phi$ is given by
\begin{align}
    \left[ \partial_t^2 + H (4 + 3 c_s^2) \partial_t + \left\{ \left( \frac{c_s k}{a} \right)^2  + 3 (c_s^2 - w) H^2 \right\} \right] \Phi_k (t) =0 \,. \label{Eq:Phi}
\end{align}
For $w=c_s^2= 1/3$, we can solve Eq.~(\ref{Eq:Phi}) analytically, obtaining 
\begin{align}
    \Phi_k(t) = 2 \zeta^p_k \, \frac{\sin y_k - y_k \cos y_k}{y_k^3}  \label{Sol:PhikRD}
\end{align}
where $\zeta^p_k$ denotes the primordial amplitude of $\zeta$ and $y_k$ is defined as
\begin{align}
    y_k (t) \equiv \frac{1}{\sqrt{3}} \frac{k}{a H} = \frac{1}{\sqrt{3}} \frac{k}{a_{\rm c} H_{\rm c}} \frac{a}{a_{\rm c}} =  \frac{1}{\sqrt{3}} \frac{k}{a_{\rm c} H_{\rm c}} \eta\,. 
\end{align}

\begin{table}[]
    \centering
    \begin{tabular}{ |c| c|c|c|c|c| } 
 \hline
  &  Self-int. & \multicolumn{3}{c|}{Int. w/ radiation }  \\[2.5pt] 
\cline{3-5}
 &  & $S_1$ (indirect) & $S_2$ (indirect) & $S_3$ (direct) \\[2.5pt] 
 \hline
Fukunaga et al. \cite{Fukunaga:2020mvq} 
& $\checkmark$ & \ding{55} & \ding{55} & \ding{55} \\[2.5pt] 
\hline
Arvanitaki et al. \cite{Arvanitaki:2019rax}  
\begin{tabular}{c}
\end{tabular} & $\checkmark$ & $\checkmark$ & $\checkmark$ & \ding{55} \\[2.5pt] 
\hline
This paper 
& $\checkmark$ & $\checkmark$ & $\checkmark$ & $\checkmark$ \\
\hline 
\end{tabular}
\caption{This table summarizes the difference among Ref.~\cite{Fukunaga:2020mvq}, Ref.~\cite{Arvanitaki:2019rax}, and this paper, which have addressed the possibility of the QCD axion clump formation, when the PQ symmetry was broken during or before inflation. 
The fluctuation of the axion can grow, being enhanced by the self-interaction, i.e., $\tilde{V}_{\tilde{\phi} \tilde{\phi}} \neq 1$ and the direct and indirect interactions with the radiation or the adiabatic perturbation $\Phi$. The contributions with $\checkmark$ are taken into account, while those with \ding{55} are ignored. 
} \label{Table:comparison}
\end{table}

In Ref.~\cite{Fukunaga:2020mvq}, the possibility of axion clump formation was addressed, focusing on the self-interaction of the QCD axion. In this analysis, the contributions of the adiabatic perturbation, $S_1,\, S_2,\, S_3$ were simply ignored (see Table \ref{Table:comparison}). In Ref.~\cite{Arvanitaki:2019rax}, considering the indirect interaction, expressed by $S_1$ and $S_2$, Arvanitaki et al. showed that $\delta \phi$ is enhanced by the adiabatic perturbation $\Phi$ around the horizon scale. In the subsequent sections, we will show that the direct interaction, expressed by $S_3$, yields a much more significant contribution at the subhorizon scales, leading to the axion clump formation under a much milder tuning of the initial condition. 

Before discussing the impact of the direct interaction, $S_3$, on the subhorizon evolution, let us point out that $S_3$ is also important in the analysis of the superhorizon evolution. In the previous subsection, we have shown that in the superhorizon limit, there is no mixing between $\sg$ and the adiabatic perturbation. Since $\sg$ is a gauge invariant variable, one can compute $\sg$ also in Newtonian gauge by evaluating the linear combination of $\delta \phi$ and $\delta \rho$, where $\delta \rho$ in Newtonian gauge is given by Eq.~(\ref{Eq:deltarho}). When we drop $S_3$ in Eq.~(\ref{Eq:KGXRD}), the mixing between $\sg$ and the adiabatic perturbation does not vanish in the superhorizon limit, resulting in the fictitious enhancement of $\sg$. Meanwhile, the superhorizon mixing between $\delta \phi$ and the adiabatic perturbation does not vanish in Newtonian gauge (see Appendix.~\ref{App:diff-gauges}).

\section{Axion clump formation: Linear analysis}  \label{Sec:Linear}
In this section, we investigate the various instability mechanisms of the QCD axion which can take place at subhorizon scales with $k/aH \leq {\cal O}(1)$. We point out two possible scenarios of QCD axion clump formation, when the PQ symmetry was broken during or before inflation. As discussed in the previous section, since $\sg$ evolves independently of the adiabatic perturbation at superhorizon scales, in this section, we discuss the evolution of $\sg$, while computing it in Newtonian gauge. Unlike at superhorizon scales, $\sg$ and $\delta \phi$ follow a similar evolution. 

\subsection{Overview of instability mechanisms}  \label{SSec:interactions}
The QCD (misalignment) axion can undergo various instability mechanisms around $T \sim T_{\rm c}$, listed below
\begin{align*}
	\text{Possible instability}
	\left\{\begin{array}{l}
	 \text{Interaction w/radiation}\left\{\begin{array}{l} \text{i) Indirect}~(S_1,\, S_2) \\
	\text{ii) Direct}~(S_3)
        \end{array}\right.\\
	\text{Self-interaction}	\left\{\begin{array}{l}\text{iii) Tachyonic instability}\\
	\text{iv) Resonance instability}
        \end{array}\right.\\
	\end{array}
	\right.
        \quad .
\end{align*}
In this subsection, we provide a brief overview of possible instability mechanisms, which can turn an almost homogeneous spatial distribution of the axion to a rather inhomogeneous one, resulting in axion clump formation. In the following, let us discuss each instability in turn.

\subsubsection{Interaction with radiation}  \label{SSSec:int_w_radiation}
The instability caused by the interaction with radiation, i) and ii), generically takes place at subhorizon scales. Since this enhancement of the inhomogeneous mode is sourced by the adiabatic perturbation, whose amplitude is typically of $10^{-5}$, it can turn even an exactly homogeneous axion distribution to an inhomogeneous one~\cite{Arvanitaki:2019rax}.

The source terms $S_1$ and $S_2$ describe the usual gravitational instability, i.e., the axion becomes inhomogeneous, falling into the potential well created by the relativistic components.
\begin{figure}
\centering
\includegraphics[width=.48\linewidth]{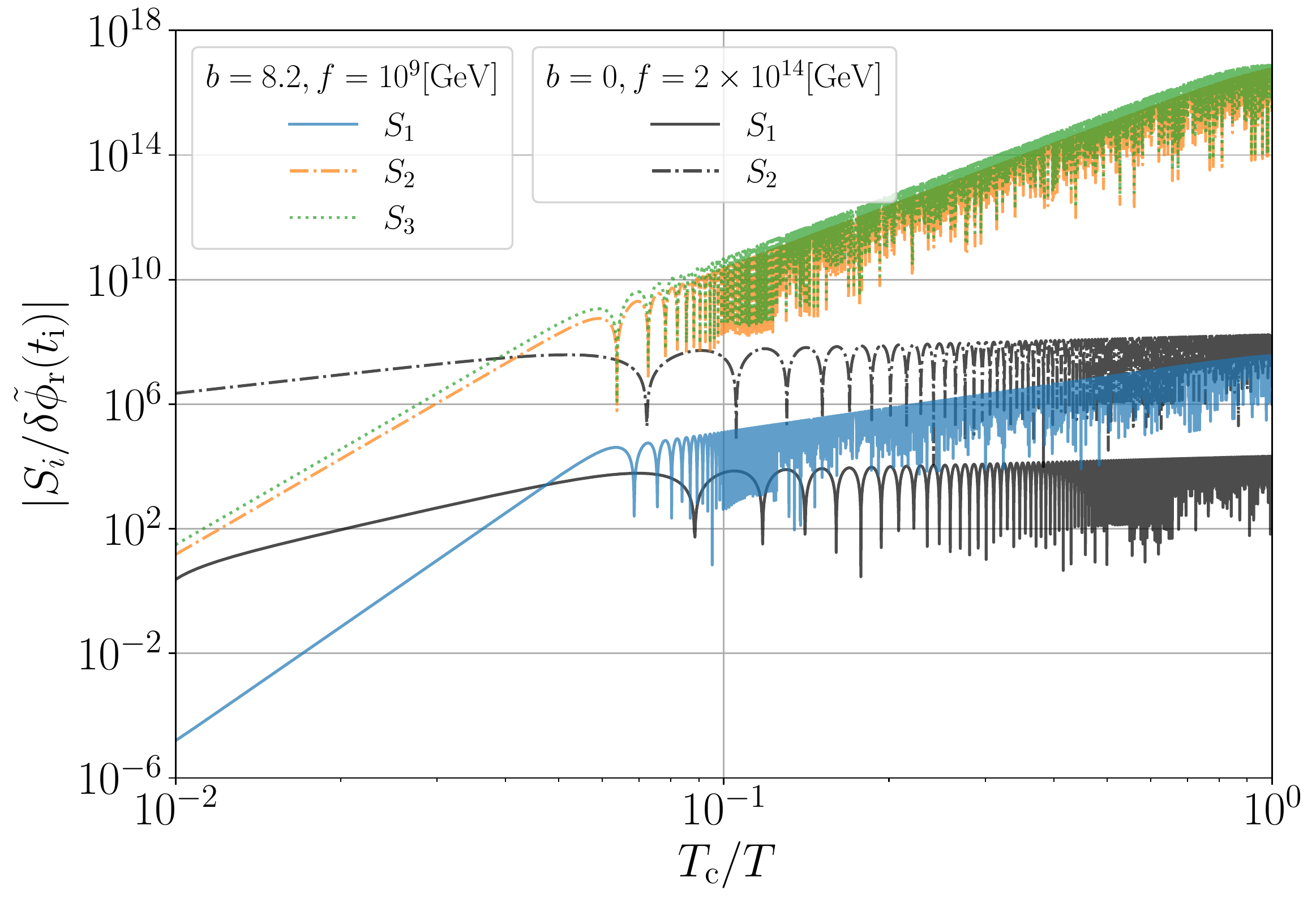}
\includegraphics[width=.48\linewidth]{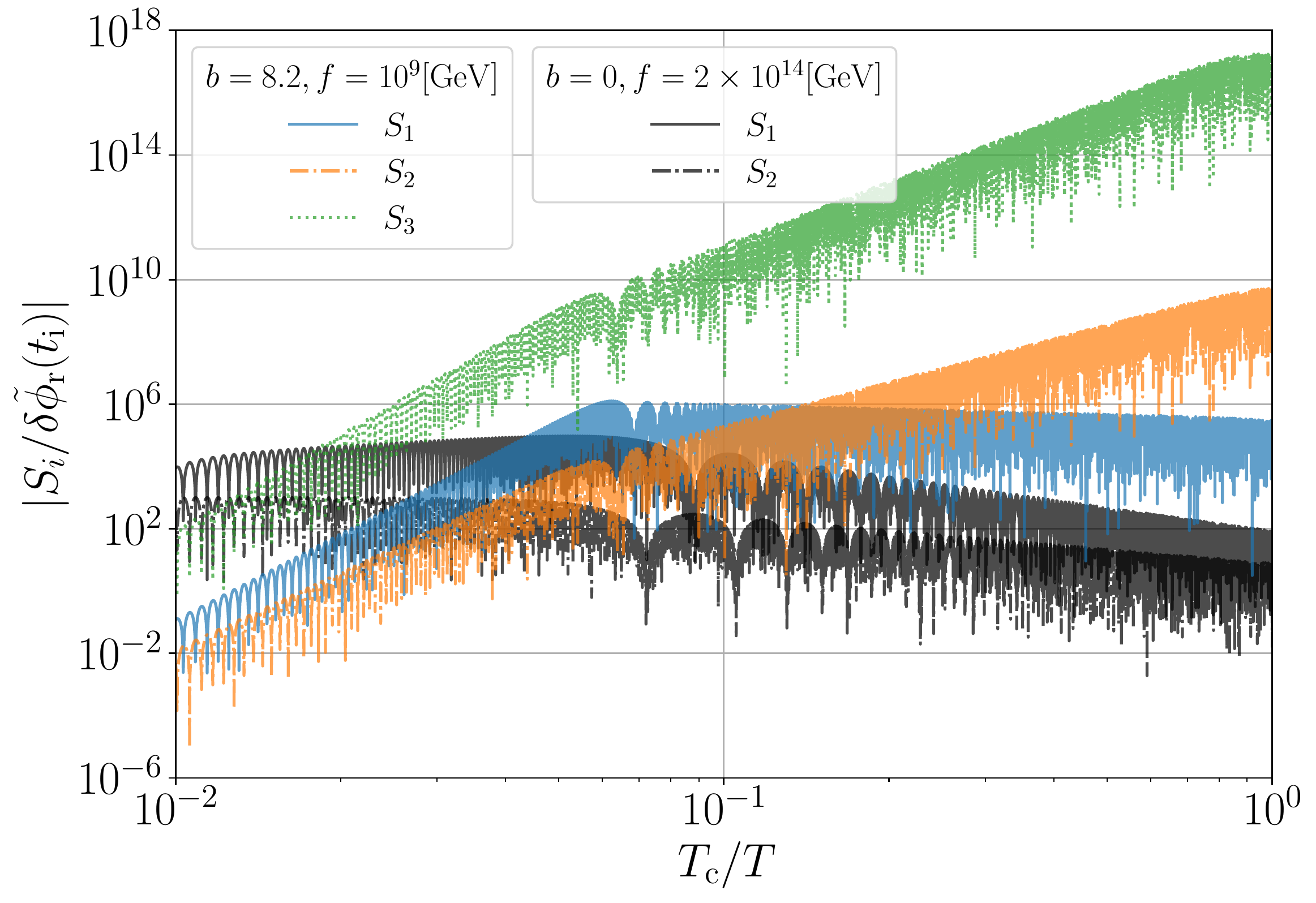}
\caption{\label{fig:3sources}These plots show the time evolution of the three source terms, defined in Eq.~(\ref{Def:sources}), which are normalized by the initial amplitude of $\sg$, $|S_i/\sg(t_{\rm i})|$, for $b=8.2$ and $\Delta = 0.9\pi$. For a comparison, we also plotted the source terms for $b=0$ by the black solid ($S_1$) and dotted ($S_2$) lines. The left panel is for $k/(a_{\rm c} H_{\rm c}) = 1$ and the right panel is for $k/(a_{\rm c} H_{\rm c}) = 10^4$.
}
\end{figure}
During radiation domination with $w=c_s^2 = 1/3$, the adiabatic curvature perturbation decays as $\Phi, d \Phi/d\eta \propto \eta^{-2}$ in the subhorizon limit (see Eq.~(\ref{Sol:PhikRD})). Using Eq.~(\ref{Exp:backgroundscaling}), which applies during the harmonic oscillation, we find that in this limit, $S_1$ and $S_2$ scale as
\begin{align}
    S_1 \propto \eta^{- \frac{10-b}{4}} \,, \qquad S_2 \propto \eta^{ \frac{3}{4} (b-2)}\,.
\end{align}
For $T> T_{\rm c}$, during which $b=8.2$, $S_1$ decays and $S_2$ increases, while for $T \leq T_{\rm c}$, both $S_1$ and $S_2$ decay in time. 

The direct interaction with radiation is described by the source term $S_3$, which scales in the subhorizon limit as
\begin{align}
    S_3 \propto \eta^{\frac{3b + 2}{4}}\,. 
\end{align}
Since $S_3$ depends on the adiabatic perturbation through the density perturbation of radiation, while $S_1$ and $S_2$ depend on it through the gravitational potential, $S_3$ grows faster than $S_2$ by $\eta^2 \propto a^2$. As a result, $S_3$ becomes the dominant source in the subhorizon limit. For $T \geq T_{\rm c}$, once the axion mass approaches to the constant value, the source term $S_3$ vanishes. Figure \ref{fig:3sources} shows the evolution of the three source terms $S_i$ ($i=1,\, 2,\, 3$). As a reference, we also show their evolution for $b=0$ by black curves.

\begin{figure}
    \centering
    \includegraphics[width=.9\linewidth]{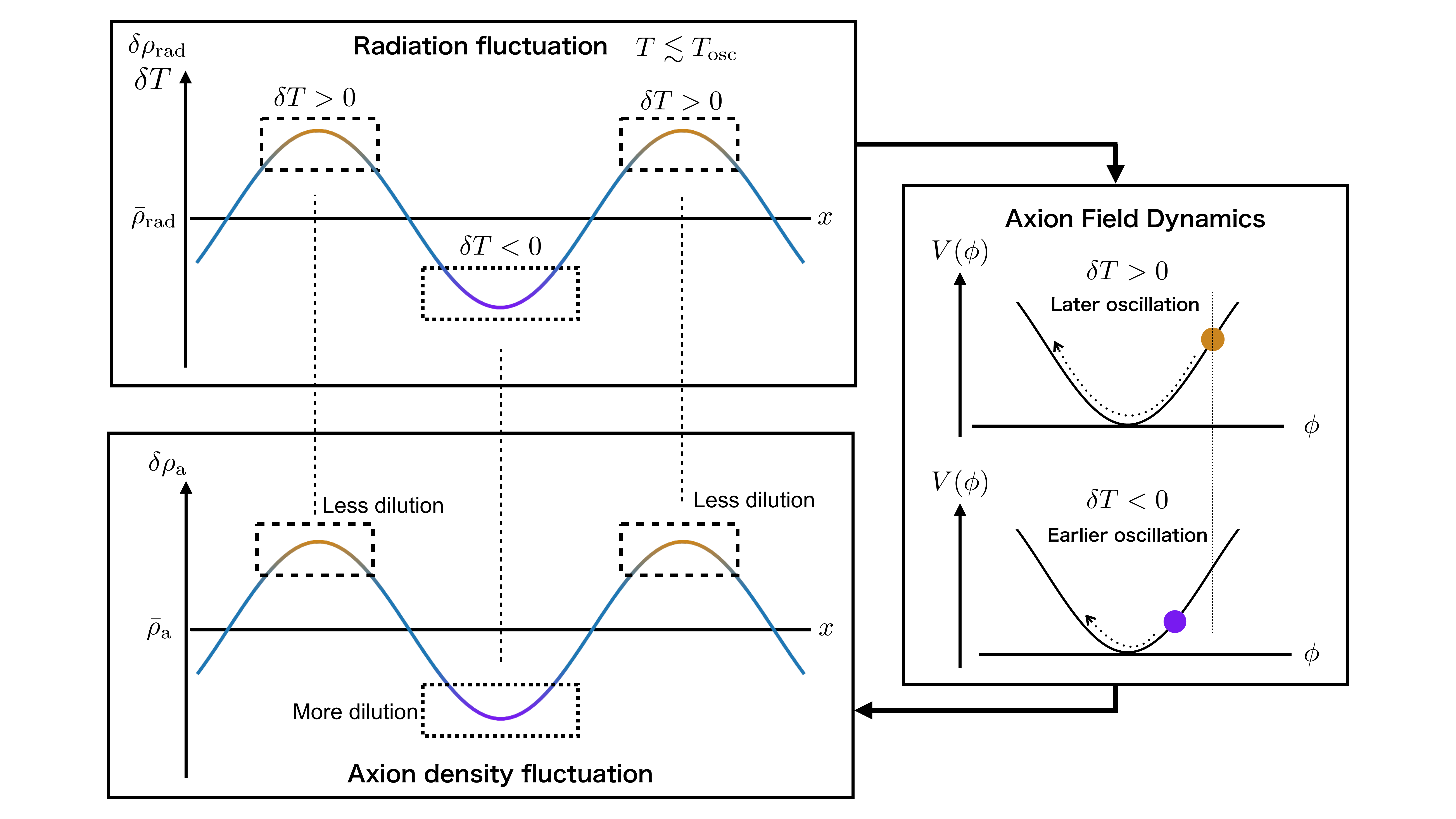}
    \caption{This figure schematically explains how the direct interaction with the radiation generates the inhomogeneity of the axion. In different spatial positions $(x)$, where the energy density or the temperature of the radiation takes different values, the mass of the axion also becomes different, experiencing different periods of dilution as a consequence of starting the oscillation at different moments. }
    \label{fig:schematic}
\end{figure}

As will be shown in the subsequent subsections, the direct interaction becomes the most significant source which enhances the inhomogeneity of the axion. This can be intuitively understood as follows (also see Fig.~\ref{fig:schematic}). In a region with a higher radiation density with a deeper potential well, the axion mass becomes smaller, roughly scaling as $m \propto T^{-b/2} \propto \rho^{- b/8}$. Since the onset of the oscillation slightly delays, the axion energy density in this region remains larger, compared to the neighborhood. This effect, described by the source term $S_3$, leads to the additional growth of the inhomogeneity, for $b> 0$. A similar mechanism for axion like particle in a completely different setup was proposed in Ref.~\cite{Hardy:2016mns}.

\subsubsection{Self-interaction}
When the initial misalignment angle is large enough, the self-interaction of the axion, quantified by $\tilde{V}_{\tilde{\phi} \tilde{\phi}} \neq 1$, also becomes important. Without the tuning of the initial condition, the self-interaction becomes negligible due to the Hubble friction, right after the onset of the oscillation. On the other hand, when the initial value of the axion is tuned around the top of the potential, iii) the tachyonic instability and subsequently iv) the resonance instability enhance the inhomogeneous modes of the axion significantly. This is because the onset of the oscillation delays under the fine-tuned initial condition~\cite{Fukunaga:2019unq}, making the influence of the cosmic expansion less significant already at the commencement of the oscillation. Then, iii) and iv) continue much longer than the case without the tuning for which $3 H_{\rm osc} \sim m_{\rm osc}$. Cosmological consequences of the delayed onset of the oscillation for axion like particles were addressed, e.g., in Refs.~\cite{Soda:2017dsu, Kitajima:2018zco, Patel:2019isj}.

\subsubsection{Previous works and prospects}
In Ref.~\cite{Fukunaga:2020mvq}, focusing on iii) and iv), while ignoring i) and ii), two of the authors addressed the possibility of the QCD axion clump formation in the scenario where the PQ symmetry was already broken during inflation. When the axion was initially around the hilltop of the potential, the inhomogeneity can grow exponentially through these two mechanisms. Without i) and ii), i.e., $S_1 = S_2 = S_3= 0$, Eq.~(\ref{Eq:KGXRD}) only has homogeneous solutions, which can be rescaled by the initial amplitude of the axion. The absence of domain walls requires that the initial amplitude of $\delta \phi$ should be smaller, when the initial position of the background axion is more tuned. As a result, for the QCD axion, the clump formation was not observed, while avoiding the domain-wall formation.

Obviously, the situation changes, when the source terms $S_i$ with $i= 1,\, 2,\, 3$ are taken into account, because the fluctuation of the radiation (or the adiabatic perturbation) can source the one of the axion (or the entropy perturbation), transforming even the exactly homogeneous initial distribution of the axion to an inhomogeneous one. In Ref.~\cite{Arvanitaki:2019rax}, taking into account the indirect interaction with QCD matters, $S_1$ and $S_2$, it was numerically shown that axion clumps with the density fluctuation of ${\cal O}(1)$ can be transiently formed under a fine-tuned initial condition with $\Delta \simeq 10^{-8}$ (see Fig.~\ref{fig:transfer_noS3} or Fig.~17 in Ref.~\cite{Arvanitaki:2019rax}). Nevertheless, the self-interaction, which has led to the formation of axion clumps, still remain important as well as after their formation, since the time scale of the cosmic expansion, which can make the self-interaction insignificant, is already longer than the time scale of the oscillation. Then, the number changing process caused by the self-interaction lets the clumps decay roughly after $10^3$ cycles of oscillation~\cite{Arvanitaki:2019rax}.  

As discussed in Sec.~\ref{SSSec:int_w_radiation}, the direct interaction with QCD matters is much more important than the indirect one. Therefore, when all the source terms are properly taken into account, axion clumps can be formed for a wider range of the initial misalignment without assuming the extreme tuning. 
\revise{Then, axion clumps may survive much longer, because the required self-interaction for their formation can be smaller. In our forthcoming paper, we will examine this speculation, using the non-linear lattice simulation. }

\subsection{Dephasing of axion through interaction with radiation}
In this subsection, computing $\sg$ by solving Eq.~(\ref{Eq:KGXRD}) numerically, we study the impact of the instability mechanisms i), ii), iii), and iv). We consider two cases, with and without the tuning of the initial condition. As discussed in the previous subsection, iii) and iv) become important only when the axion was tuned around the potential maximum.

\subsubsection{Without tuning of initial condition}
\begin{figure}
\centering
\includegraphics[width=.48\linewidth]{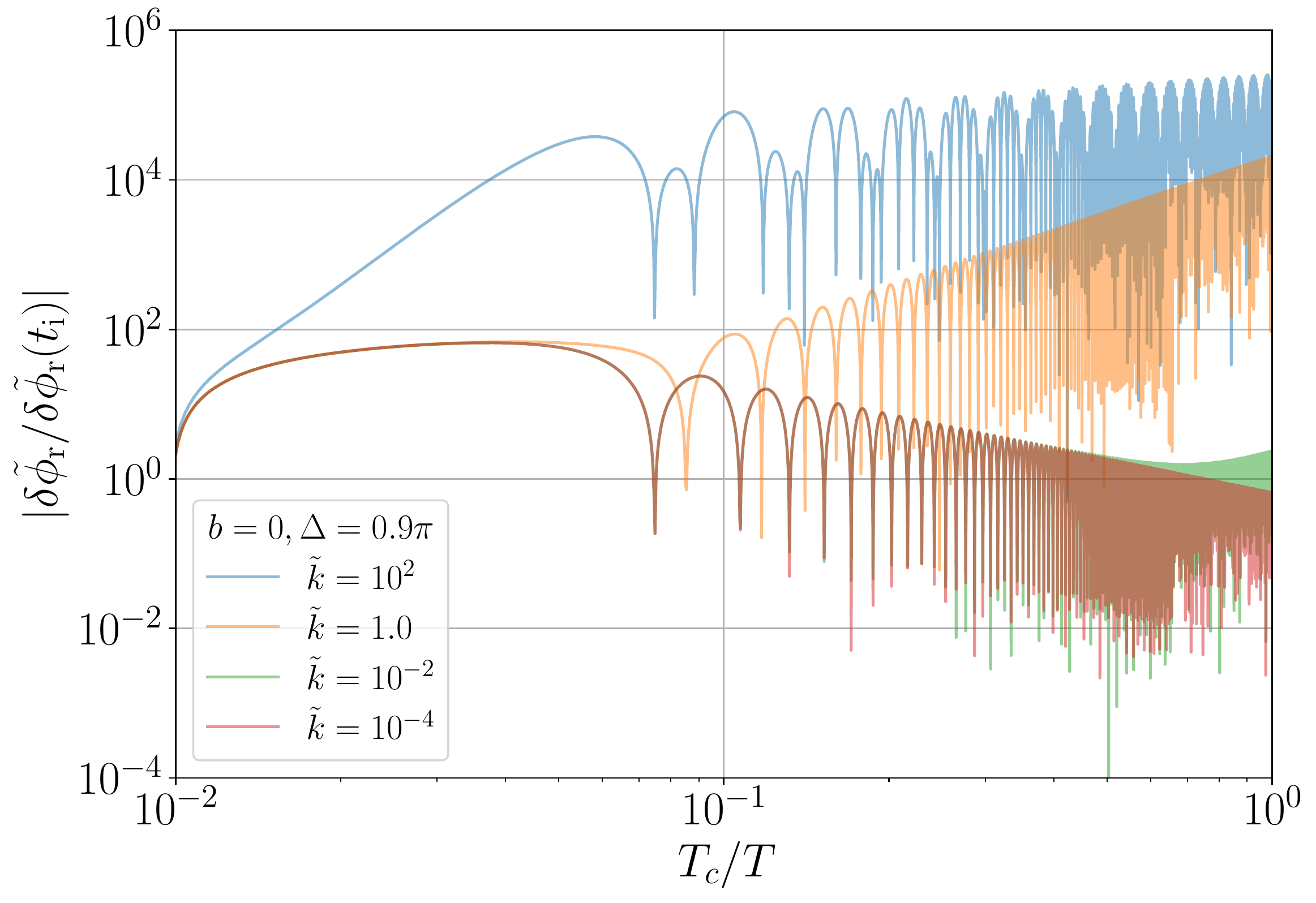}
\includegraphics[width=.48\linewidth]{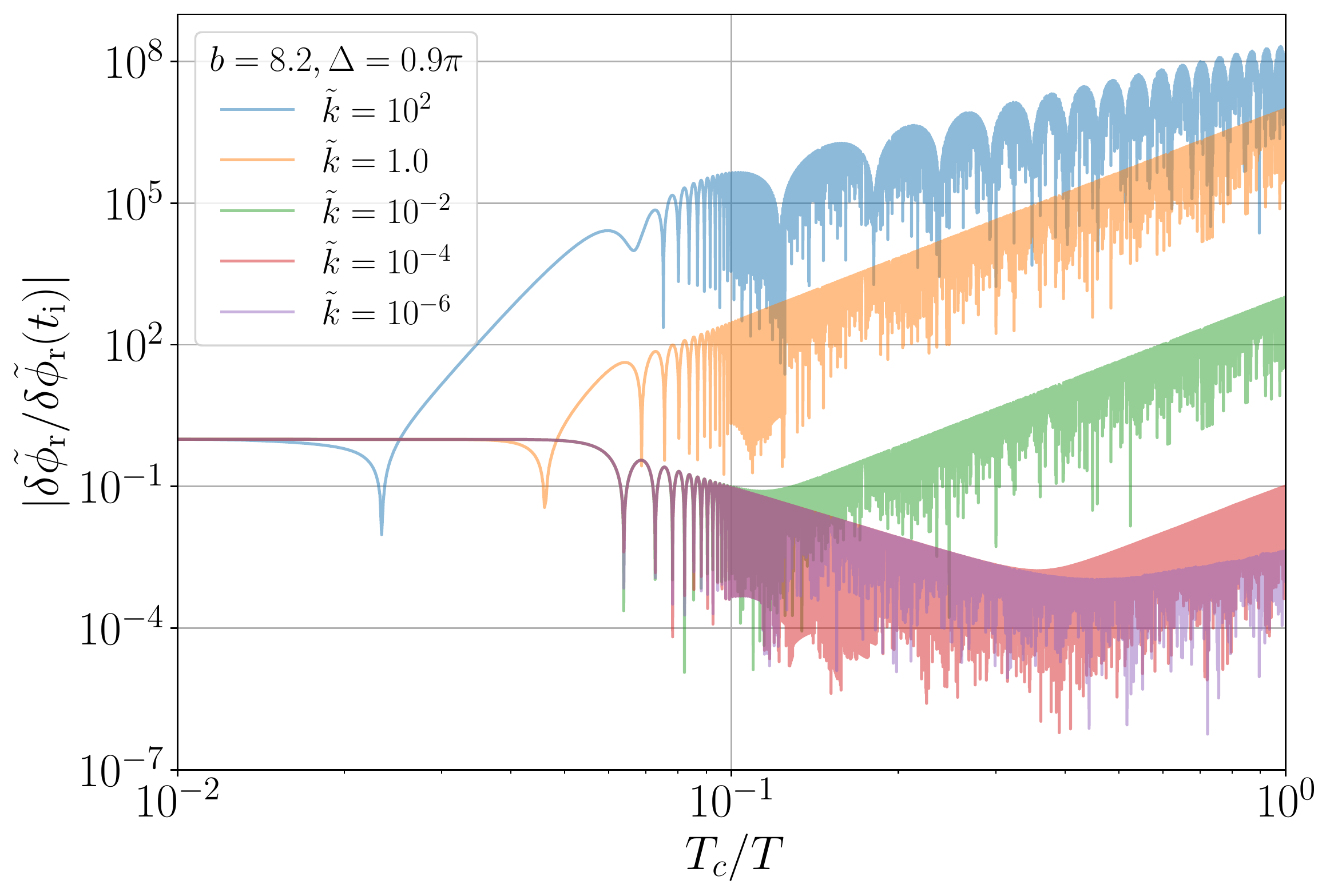}
\caption{These plots show the evolution of $|\delta \tilde{\phi}_{\rm r}/\delta \tilde{\phi}_{\rm r}(t_{\rm i})| = |\delta \tilde{\phi}_{\rm r}/(r_{\rm axi}\zeta^{\rm p})|$ for $\Delta = 0.9\pi$ and $r_{\rm axi} = 10^{-5}$. The left panel is for $b=0$ and $f=2\times10^{14}$ GeV, and the right one is for $b=8.2$ and $f=10^{9}$ GeV. The wavenumbers are $\tilde{k}= k/(a_{\rm c} H_{\rm c}) = 10^{-6}$ (purple), $10^{-4}$ (red), $10^{-2}$ (green), $1$ (orange), and $10^2$ (blue). Here, we do not show it, but the amplitude of $k/(a_{\rm c} H_{\rm c}) = 10^4$ also follows a similar evolution to the one for $k/(a_{\rm c} H_{\rm c}) = 10^2$, while it oscillates much more rapidly. In this case, at around $T=T_{\rm c}$, a wide range of wavenumbers with $k/(a_{\rm c} H_{\rm c}) \geq 10^{-4}$ are enhanced by the gravitational instability due to the adiabatic mode. \label{Fig:comparisonb}}
\end{figure}
Figure \ref{Fig:comparisonb} shows the evolution of the transfer function $\sg/\sg(t_{\rm i})$ for various wavenumbers $\tilde{k} = k/(a_{\rm c} H_{\rm c})$ for $b=0$ (left) and for $b= 8.2$ (right). To highlight the significant enhancement of $\sg$ due to the direct interaction, described by $S_3$, we also show the evolution for $b=0$, in which $S_3$ vanishes because of the absence of the direct interaction. In this case, the indirect interaction i), described by $S_1$ and $S_2$, is the only source of the instability. For $b=0$, We choose the PQ scale $f = 2\times 10^{14}\,{\rm GeV}$ so that the oscillation starts roughly at the same temperature as $f = 10^9\, {\rm GeV}$ and $b=8.2$. The initial amplitude of the fluctuation of the axion is set to $r_{\rm axi}= 10^{-5}$, where $r_{\rm axi}$ is defined as
\begin{align}
    r_{\rm axi} \equiv |\delta\tilde{\phi}_{\rm r}(t_{\rm i},\, k)/\zeta^{\rm p}(k)|.
\end{align}
As long as the initial amplitude of the fluctuation of the axion is smaller than the one of the radiation, the evolution of $\sg$ becomes independent of $r_{\rm axi}$, once the mixing between the axion and the radiation sets in.

Both for $b=0$ and $b=8.2$, the perturbation of the axion in a wide range of the wavenumbers $ k/(a_{\rm c} H_{\rm c}) \geq {\cal O}(10^{-4})$ is enhanced, being sourced by the adiabatic perturbation. Compared to the left panel of Fig.~\ref{Fig:comparisonb} with $b=0$, the right panel with $b=8.2$ shows much more significant enhancement due to the impact of the direct interaction, which dominates the one of the indirect interactions at the subhorizon scales. This enhancement sets in after the axion starts to acquire the potential through the interaction with QCD matters, because the mixing with the adiabatic perturbation vanishes in the high temperature limit, where $\dot{\phi}$ and $V_\phi$ are still negligibly small.

The enhancement through the interaction with the radiation becomes more significant for a smaller PQ scale $f$, for which $\sg$ is sourced by the adiabatic perturbation during a longer period. When there is no tuning of the initial condition, e.g., for $\Delta = 0.9\pi$, the misalignment angle of the QCD axion generically acquires the ${\cal O}(1)$ fluctuation at $T \sim T_{\rm c}$ for $f \lesssim 3\times10^9$ GeV, being sourced by the adiabatic perturbation whose amplitude is ${\cal P}_\zeta = 2.1\times10^{-9}$, which is the Planck fiducial value \cite{Akrami:2018odb}.

\begin{figure}
    \centering
    \includegraphics[width=0.49\linewidth]{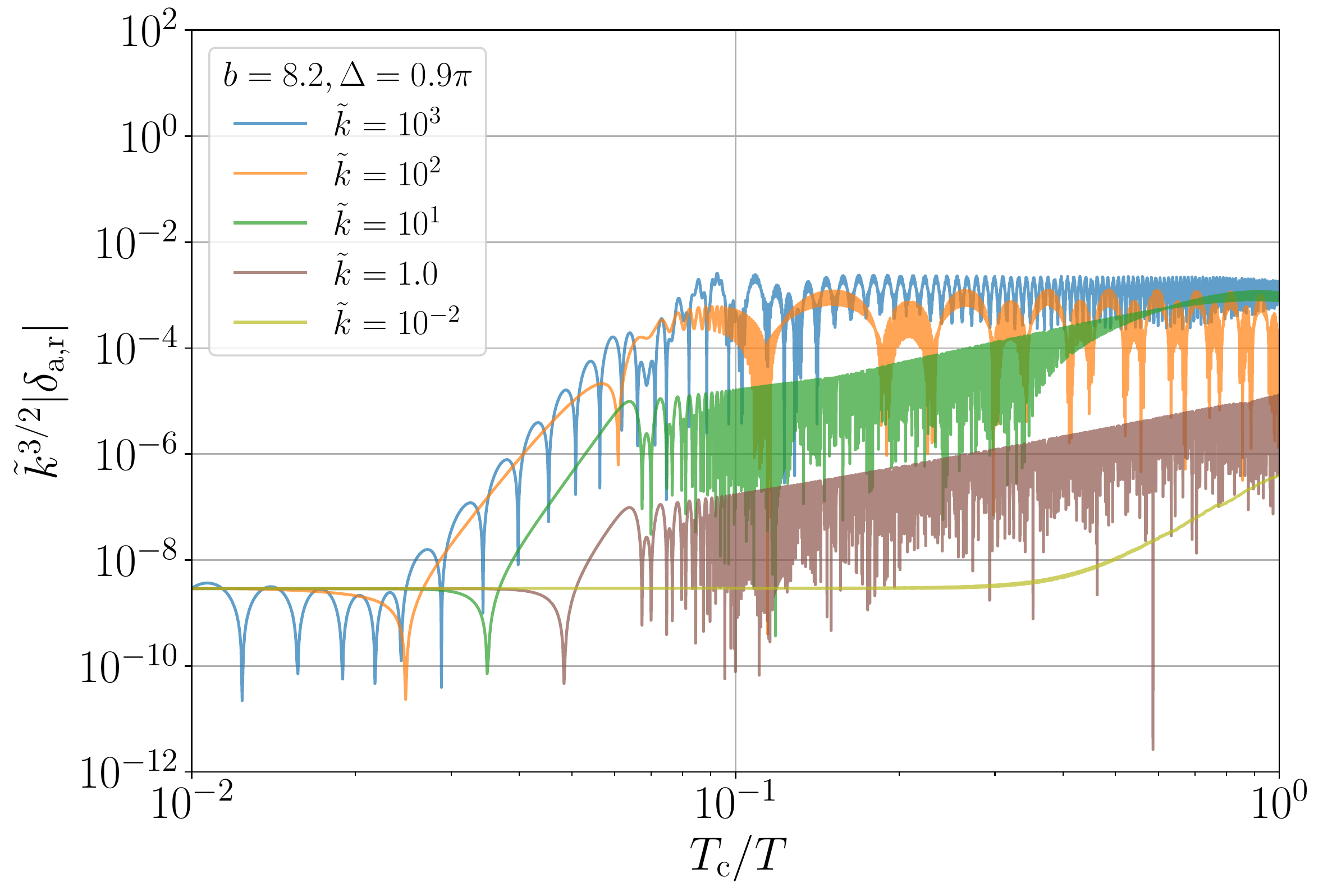}
    \includegraphics[width=0.49\linewidth]{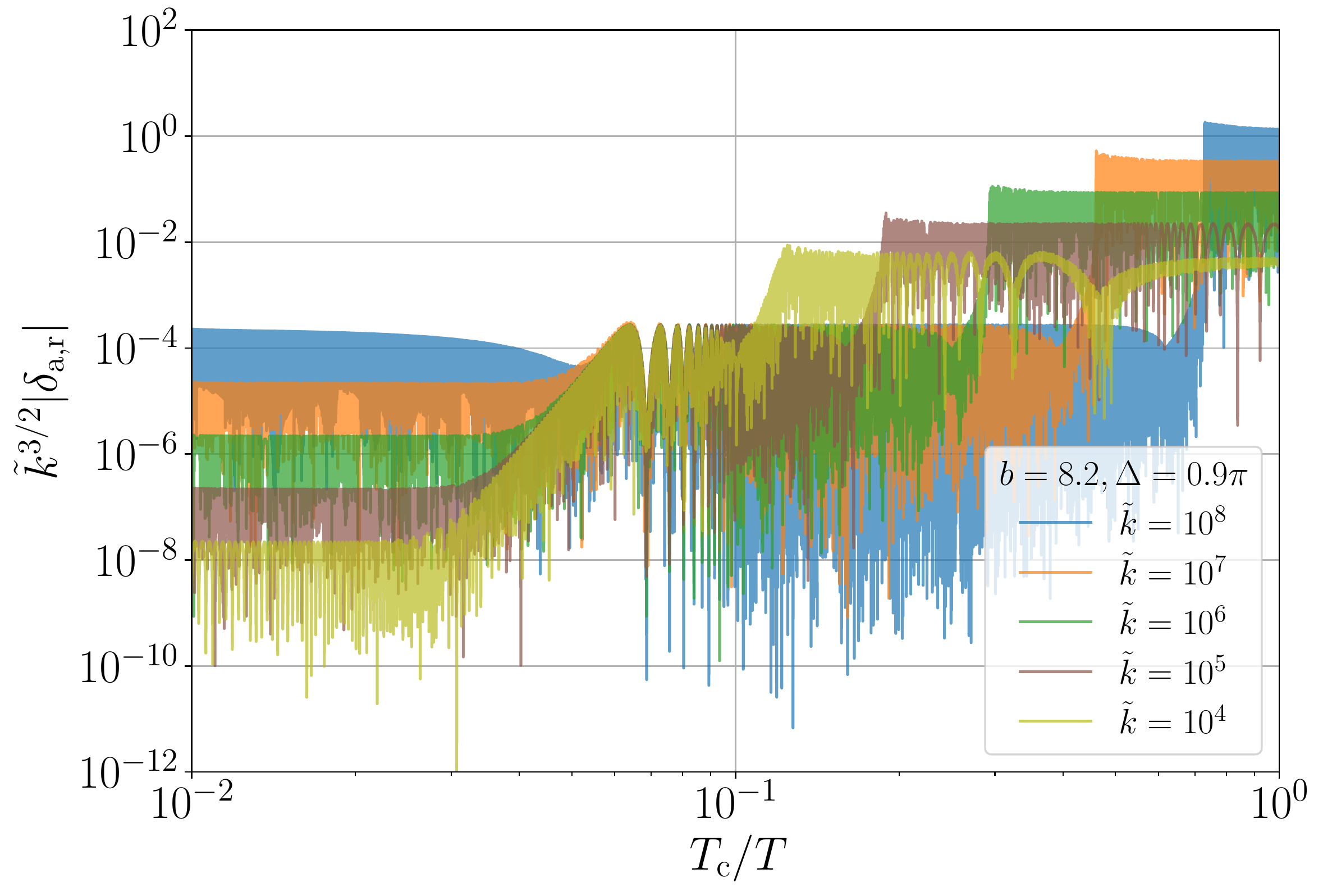}
    \caption{These plots show the time evolution of the axion energy density perturbation evaluated on the uniform radiation slicing for $b=8.2, r_{\rm axi}=10^{-5}, \Delta = 0.9\pi, f=10^9\,{\rm GeV}$.}  \label{Fig:deltarho_wo_tuning}
\end{figure}

Figure \ref{Fig:deltarho_wo_tuning} shows the time evolution of the density perturbation of the axion on the uniform radiation density slicing, $\delta \rho =0$\footnote{The entropy perturbation is usually characterized by using the (relative) entropy perturbation for the components $\alpha$ and $\beta$, defined as~\cite{Malik:2002jb}
\begin{align}
   S_{\alpha \beta} \equiv  3 \left( \zeta^{\alpha} -  \zeta^{\beta} \right) = - 3 H \left( \frac{\delta \rho^{\alpha}}{\dot{\rho}^{\alpha}} - \frac{\delta \rho^{\beta}}{\dot{\rho}^{\beta}}\right)\,, \label{Def:S}
\end{align}
where $\zeta^\alpha$ and $\zeta^\beta$ denote the curvature perturbation ${\cal R}$ on the time slicings $\delta \rho_\alpha=0$ and $\delta \rho_\beta=0$, respectively. Using Eq.~(\ref{Exp:drhophi_UD}), the relative entropy between the axion and the radiation, $S$, can be given by
\begin{align}
     S = - \frac{3H}{\dot{\rho}_{\rm a}} \delta \rho_{{\rm a,r}}\,.
\end{align}
Especially, once the interaction between the axion and the radiation is turned off, satisfying the energy conservation independently as $\dot{\rho}_{\rm a} = - 3 H \rho_{\rm a}$, $S$ agrees with the density perturbation of the axion in the uniform radiation slicing, i.e., $S = \delta \rho_{{\rm a,r}}/\rho_{\rm a}=\delta_{{\rm a}, {\rm r}}$. }, 
\begin{align}
    \delta_{{\rm a}, {\rm r}} \equiv \frac{\delta \rho_{{\rm a}, {\rm r}}}{\rho_{\rm a}}\,, 
\end{align}
where $\delta \rho_{{\rm a}, {\rm r}}$ is given by Eq.~(\ref{Exp:drhophi_UD}). Although the inhomogeneity in the phase of the axion grows, developing the ${\cal O}(1)$ fluctuation during $T > T_{\rm c}$, the fluctuation of the axion energy density stops growing. This can be confirmed, e,g, by comparing the blue curve in the right panel of Fig.~\ref{Fig:comparisonb} to the orange one in the left panel of Fig.~\ref{Fig:deltarho_wo_tuning}. The final amplitude of $\delta \rho_{{\rm a,r}}/\rho$ becomes larger for larger wavenumbers. 

The different time evolution between the phase $\delta \phi_{\rm r}$ and the energy density $\delta \rho_{{\rm a,r}}$ can be understood as follows. As shown in Eq.~(\ref{Exp:drhophi_UD}), $\delta \rho_{{\rm a,r}}$ is given by the summation of the kinetic energy, the potential energy, and the effect of the gravitational potential. Although the interaction with the radiation causes the dephasing of the axion until $T \sim T_{\rm c}$, enhancing $\delta \phi_{\rm r}$, this effect is canceled out between the kinetic energy and the potential energy in the limit $k/(am) \ll 1$ and $H \ll m$. Therefore, once $k/(am) \propto a^{-b/2-1}$ becomes ${\cal O}(1)$, the energy density stops growing as shown in Fig.~\ref{Fig:deltarho_wo_tuning}. This cancellation can be schematically understood by solving
\begin{align}
    \delta \ddot{\phi}_k + m^2 \delta \phi_k \simeq \phi \times {\cal A}
\end{align}
which approximates the KG equation in Newtonian gauge in the limit $H/m  \ll 1$ and $k/(am) \ll 1$ and in the absence of the self-interaction. In the source term, we have factored out the oscillating contribution in the time scale of $1/m$, leaving ${\cal A} \propto \delta \rho/\rho$, which varies much more slowly in the limit $k/(am) \ll 1$ and $H \ll m$. The sourced solution of this equation is given by
\begin{align}
   \delta \phi_k (t) = - \int^{t} \frac{d t'}{2 m} i (e^{i m(t- t')} - {\rm c.c.}) \phi(t') {\cal A}(t') \,. 
\end{align}
Expressing the background solution as $\phi(t) \propto \cos(mt + \psi)$ with $\psi$ being a constant phase shift and dropping the high-frequency contribution, we obtain $\delta \phi_k (t) \propto {\cal A} \sin (mt + \psi)$. Therefore, the growing contribution is cancelled by summing the kinetic energy and the potential energy as 
$\dot{\phi} \delta \dot{\phi}_{\rm r} + m^2 \phi \delta \phi_{\rm r} = \dot{\phi} \delta \dot{\phi} + m^2 \phi \delta \phi + \cdots$, where we have abbreviated the terms which are proportional to the fluctuation of the radiation energy density, $\delta \rho/\rho$. Therefore, the dephasing of the axion due to the interaction with the radiation does not result in the growth of the inhomogeneity in the energy density of the axion for $k/(am) \ll 1$. The smaller $k$ mode stops growing earlier, when it becomes $k/(am) \sim {\cal O}(1)$, where $am$ grows as $am \propto a^{1+ b/2}$. This cancellation happens both in the Newtonian gauge and the uniform radiation density slicing.
 
The high $k$ modes in the right panel of Fig.~\ref{Fig:deltarho_wo_tuning} grow with two steps. The first one, which takes place around the onset of the oscillation, can be understood as explained in Fig.~\ref{fig:schematic}. Meanwhile, the second one is due to the resonance caused by the oscillating (external) source term, $S_3$. The second growth also remains in the density perturbation without being canceled out. When we manually turn off the oscillating phase in $S_3$, as is expected, the first growth remains almost the same, while the second growth disappears. The first one generates the fluctuation of the axion, whose amplitude is slightly larger than the one of the adiabatic perturbation, while the second one can further enhance the fluctuation.  

The impact of the resonance was also addressed by Sikivie and Xue in Ref.~\cite{Sikivie:2021trt}. However, their resultant spectrum is quite different from ours, e.g., $\Delta=0.9\pi$ in Fig.~\ref{Fg:specrum}. \revise{This is mainly because, instead of the density perturbation of the axion, whose evolution was just addressed above,} they computed $\delta \rho_{\rm SX} \sim (m \delta \phi)^2$, which is expressed only in terms of the potential energy. Because of that, the cancellation between the kinetic and potential terms does not occur for low $k$ modes in their computation. Consequently, they found that the excited fraction of the axion, which is inferred from $\delta \rho_{\rm SX}$, can reach ${\cal O}(1)$ in the low $k$ region.

\subsubsection{With tuning of initial condition}
\begin{figure} 
\centering
\includegraphics[width=.48\linewidth]{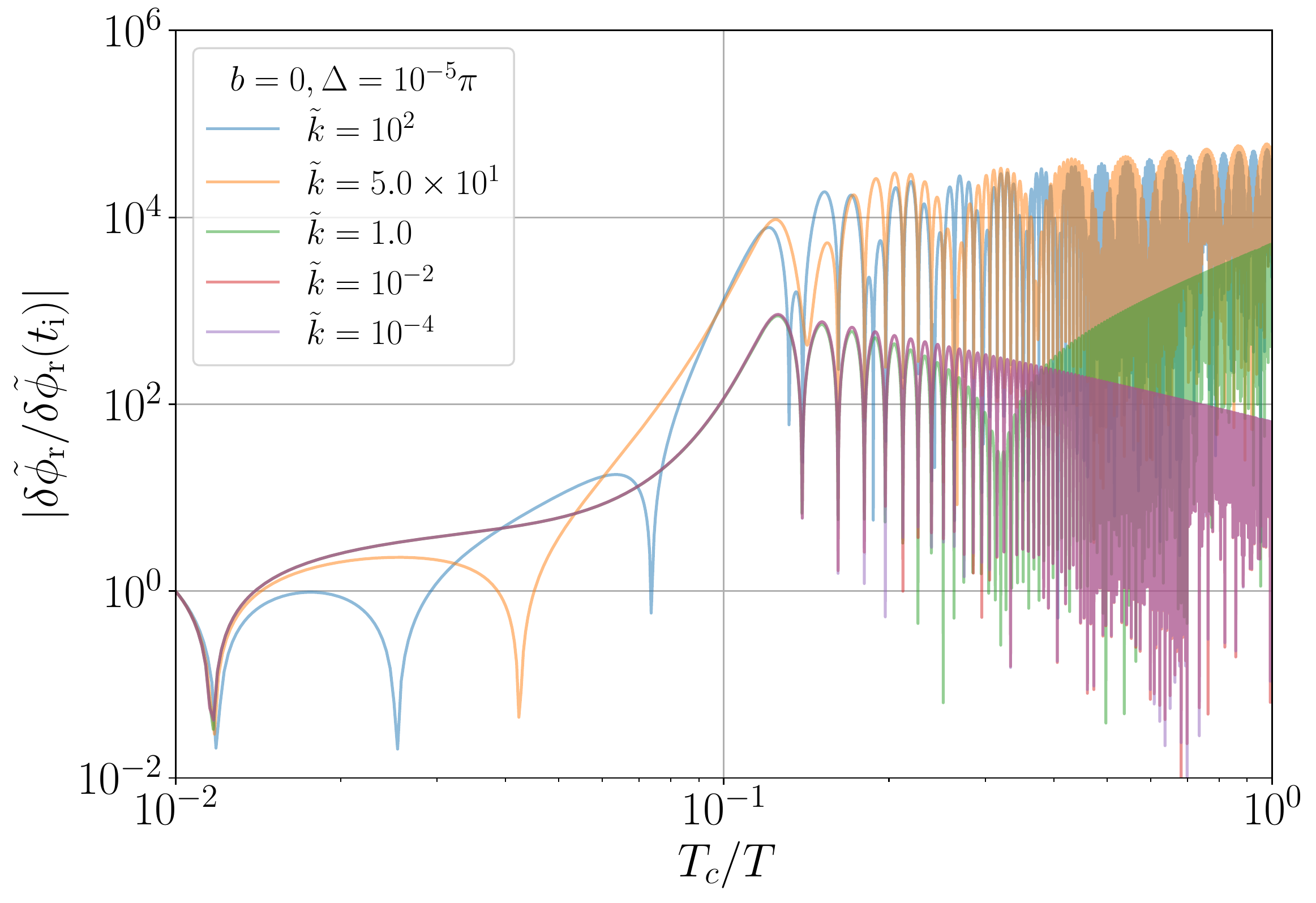}
\includegraphics[width=.48\linewidth]{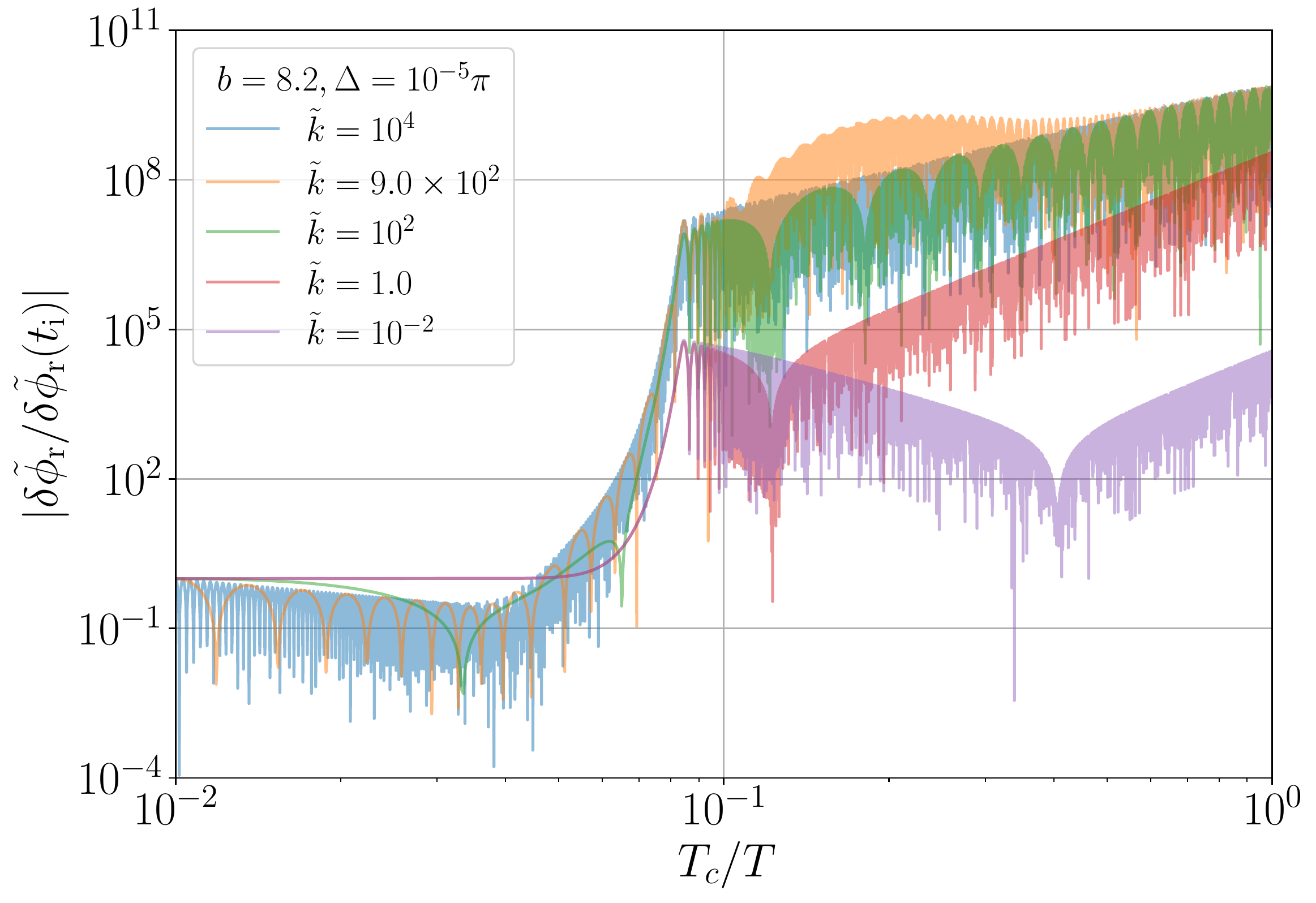}
\caption{These plots are the same as Fig.~\ref{Fig:comparisonb} except that the initial condition of the axion is tuned as $\Delta=10^{-5}\pi$. The wavenumbers $k/a_{\rm c}H_{\rm c} = 50$ for $b=0$ (the left panel) and $k/a_{\rm c}H_{\rm c} = 9\times10^2$ for $b=8.2$ (the right panel) are enhanced by the resonance instability due to the self-interaction. \label{Fig:comparisonbwithtuning}}
\end{figure}
\begin{figure}
    \centering
    \includegraphics[width=0.49\linewidth]{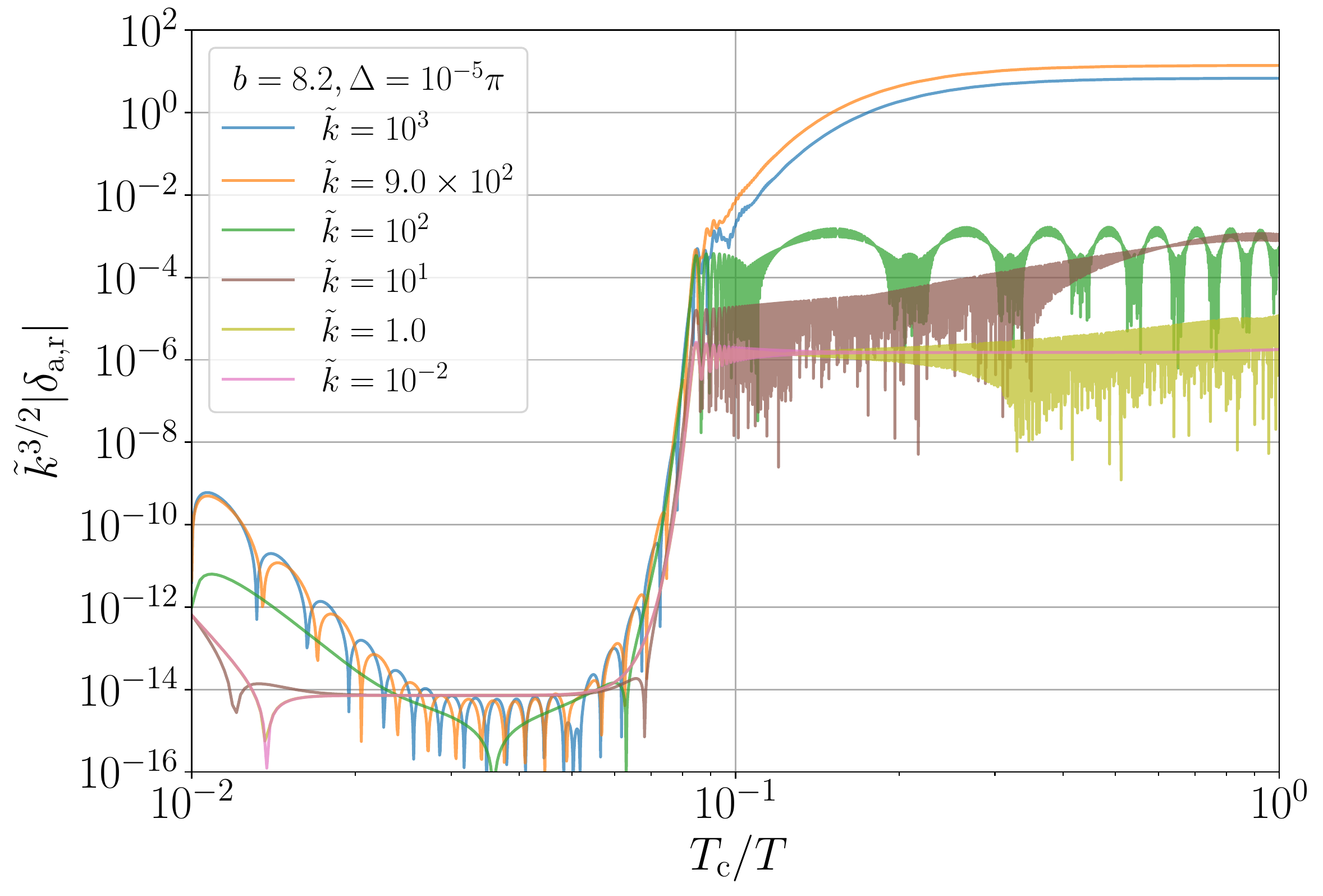}
    \includegraphics[width=0.49\linewidth]{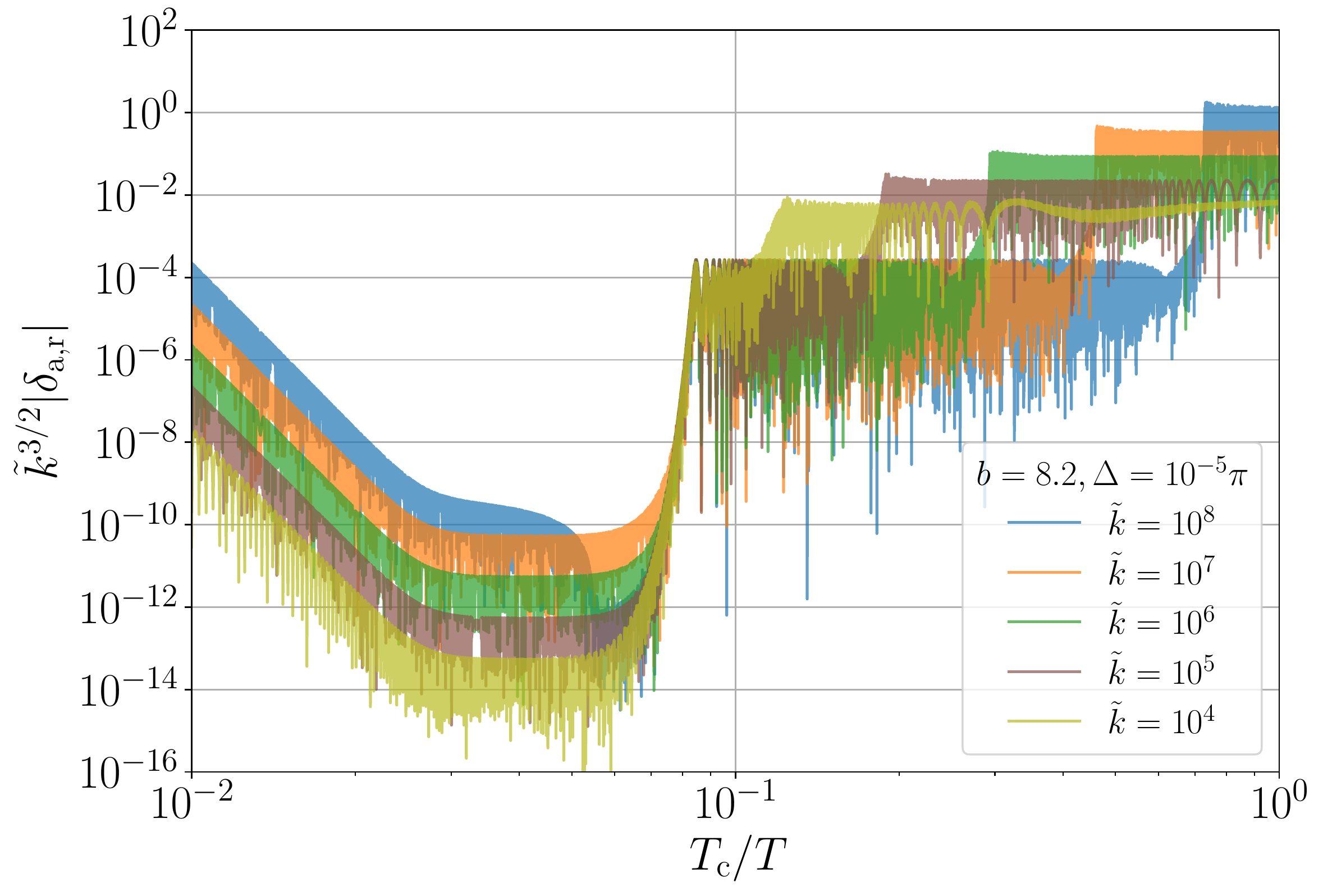}
    \caption{These plots show the time evolution of the axion density perturbation for $b=8.2, r_{\rm axi}=10^{-5}, \Delta = 10^{-5}\pi$ and $f=10^9\,{\rm GeV}$.} \label{Fig:deltarhoawithtuning}
\end{figure}
Figure \ref{Fig:comparisonbwithtuning} shows the time evolution of the transfer function for $\delta \phi$, when the axion was initially tuned around the top of the potential. In this case, in addition to the interaction with the radiation, the tachyonic instability iii) and the resonance instability iv) due to the self-interaction also can enhance the fluctuation of the axion. In the right panel of Fig.~\ref{Fig:comparisonbwithtuning} with $b=8.2$, just before the onset of the oscillation, the superhorizon modes $k/(a_{\rm c} H_{\rm c}) = 10^{-2},\, 1.0$ are enhanced due to the tachyonic instability and the subhorizon modes $k/(a_{\rm c} H_{\rm c}) = 10^2,\, 9.0 \times 10^2,\, 10^4$ are enhanced by being sourced by the fluctuation of the radiation. Right after the oscillation commences, $k/(a_{\rm c} H_{\rm c}) = 9.0\times10^2$ is enhanced by the resonance instability due to the self-interaction. This mode roughly corresponds to $k/(am) = {\cal O}(1)$, when the resonance took place. The impact of the self-interaction was discussed in Refs.~\cite{Arvanitaki:2019rax, Fukunaga:2020mvq}. As shown in Fig.~\ref{Fig:deltarhoawithtuning}, the growth due to the resonance instability can be also seen in the fluctuation of the axion density without being canceled out.

\subsection{New scenario of axion clump formation}
\begin{figure}  
    \centering
    \includegraphics[width=0.49\linewidth]{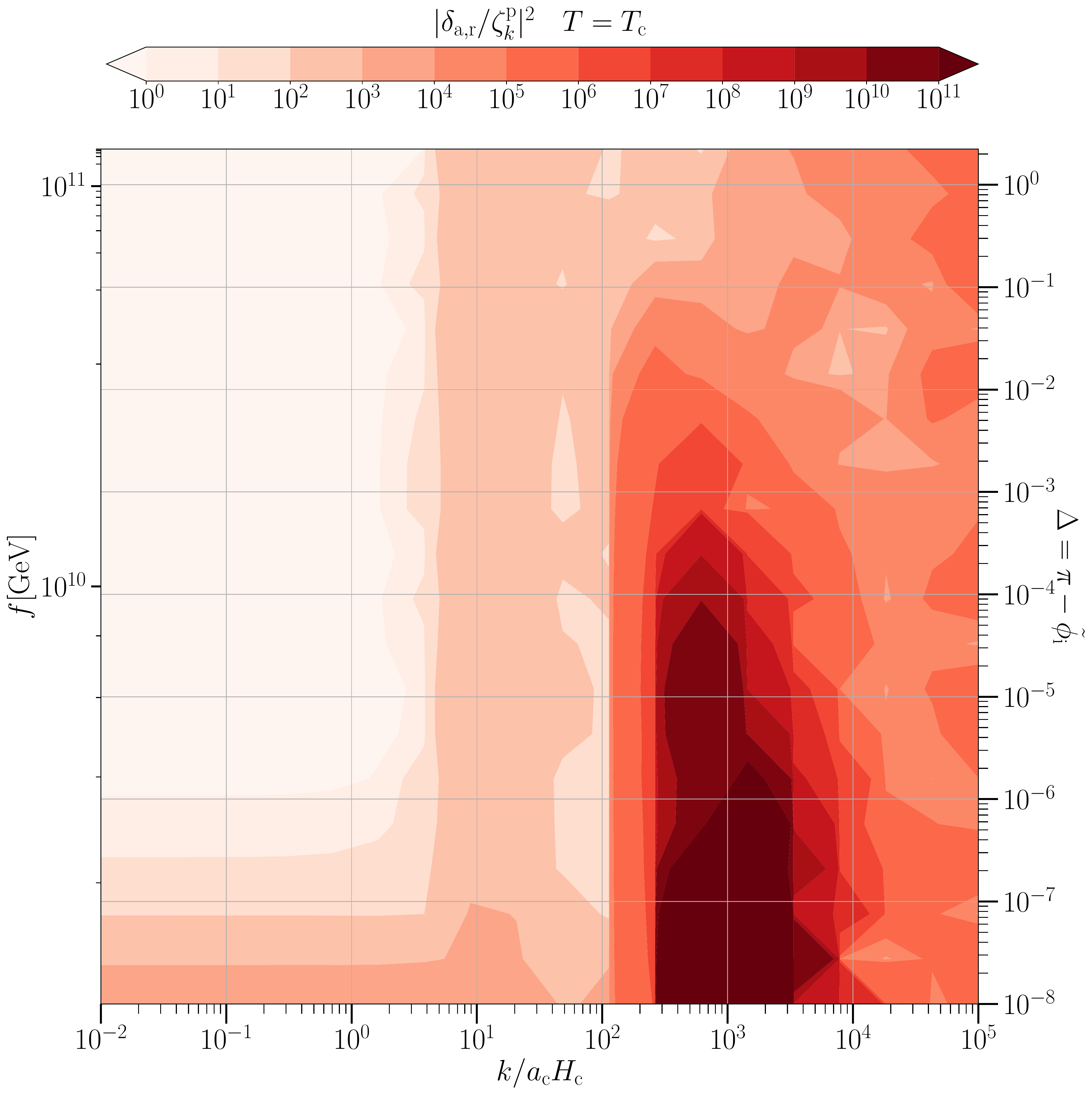}
    \caption{This plot shows the transfer function of the axion density perturbation for $b=8.2, r_{\rm axi} = 10^{-5}$ at $\eta=1$, i.e., $T=T_{\rm c}$. The decay constant $f$ and $\Delta$ are chosen so that they satisfy Eq.~(\ref{eq:f-delta}), which is obtained requiring $\Omega_{\rm a} = \Omega_{\rm DM} = 0.23$. \label{Fg:transferfn}}
\end{figure}
\begin{figure} 
    \centering
    \includegraphics[width=0.7\linewidth]{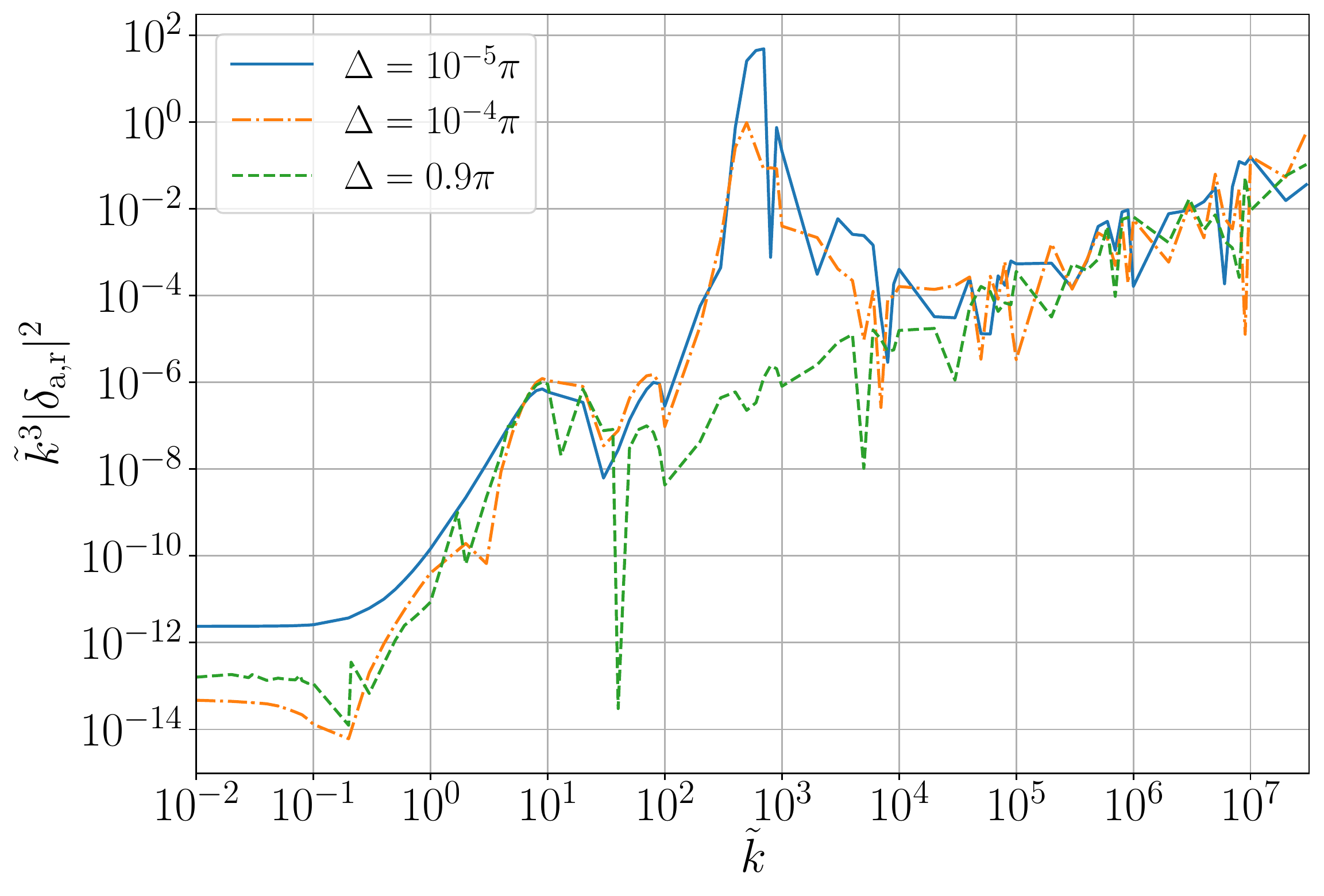}
    \caption{The spectrum of the axion density perturbation for $b=8.2, r_{\rm axi} = 10^{-5}$ at $\eta=1$ or equivalently $T= T_{\rm c}$. 
    For $\Delta= 10^{-4} \pi,\, 10^{-5} \pi$, the decay constant $f$ is determined by imposing Eq.~(\ref{eq:f-delta}), so that the QCD axion saturates the total dark matter abundance. The green dashed line corresponds to $\Delta=0.9\pi$ at $f=10^9\,{\rm GeV}$. The primordial amplitude of the adiabatic perturbation is set to ${\cal P}_\zeta = 2.1\times 10^{-9}$ \cite{Planck:2018vyg}. Once the amplitude reaches ${\cal O}(1)$, we cannot trust the result of the linear analysis.  \label{Fg:specrum} }
\end{figure}

Figure \ref{Fg:transferfn} shows the transfer function of the axion density perturbation in the uniform radiation slicing, $\delta_{{\rm a}, {\rm r}} \equiv \delta \rho_{{\rm a}, {\rm r}}/\rho_{\rm a}$, evaluated at $T=T_{\rm c}$ for different values of $f$ or $\Delta$, which are related by requiring that the QCD axion saturates the total dark matter abundance. As has been discussed in this section, the fluctuation of the axion $\delta_{{\rm a}, {\rm r}}$ is enhanced only through the interaction with the radiation for $\Delta = {\cal O}(1)$, while it is enhanced by both the interaction with the radiation and the self-interaction for $\Delta \ll 1$. For the former, $\delta_{{\rm a}, {\rm r}}$ grows more in smaller scales, since the interaction with the radiation becomes important only after the horizon crossing. For $\Delta \ll 1$, in addition to this enhancement, the resonance instability significantly enhances the modes around $k/(a_{\rm c} H_{\rm c}) \sim 7 \times 10^2$, creating the spectral peak. Figure \ref{Fg:specrum} shows the spectrum of $\delta_{{\rm a}, {\rm r}}$ for $\Delta = 0.9\pi$ without the initial tuning and for $\Delta = 10^{-4} \pi,\, 10^{-5} \pi$ with the tuning. The primordial amplitude of the adiabatic perturbation $\zeta_k^{\rm p}$ is set to ${\cal P}_\zeta = 2.1\times 10^{-9}$ \cite{Planck:2018vyg}. We find the resonance peak due to the self-interaction for $\Delta = 10^{-4} \pi,\, 10^{-5} \pi$ on top of the enhancement due to the interaction with the fluctuation of the radiation, which becomes more prominent in the small scale limit.  

The QCD axion clumps whose density contrast is ${\cal O}(1)$ can be formed in two different scenarios, even if the PQ symmetry was already broken during inflation. First, when the primordial amplitude is ${\cal P}_\zeta \sim 10^{-9}$ around the horizon scale at $T \sim T_{\rm c}$, i.e., $k \simeq a_{\rm c} H_{\rm c}$, as a simple extrapolation from the CMB scale, the density perturbation of the axion can reach ${\cal O}(1)$, forming the axion clumps, when the initial condition is tuned as $\Delta < {\cal O}(10^{-3})$. Second, even without the initial tuning, being $\Delta = {\cal O}(1)$, the axion clumps can be formed, when the primordial amplitude at $k \simeq a_{\rm c} H_{\rm c}$ is enhanced, being ${\cal P}_\zeta \geq {\cal O} (10^{-3})$, e.g., as a consequence of a non-trivial time evolution during inflation. As shown in Fig.~\ref{Fg:specrum}, $\delta_{{\rm a}, {\rm r}}$ can reach ${\cal O}(1)$ in the deep subhorizon scales as well as for ${\cal P}_\zeta \sim 10^{-9}$ and without the initial tuning. However, this enhancement will not lead to a formation of axion miniclusters, being smoothed out by the pressure, since the wavenumbers of these modes are larger than the Jeans scale $k_{\rm J}$, given by
\begin{align}
    \frac{k_{\rm J}}{a_0} \simeq 6.7 \times 10^3 \left( \frac{a}{a_0} \right)^{1/4} \left( \frac{m_{\rm a}}{\mu {\rm eV}} \right)^{1/2} {\rm pc}^{-1} 
\end{align}
for $\Omega_{\rm a} h^2 = 0.12$.

\section{Discussions}\label{sec:conc}
In this paper, we studied the impact of the direct interaction between the axion and the QCD matters. As is shown in Sec.~\ref{Sec:Linear} and also in Appendix \ref{App}, the direct interaction is crucial both in the superhorizon and subhorizon scales. While the direct interaction keeps on enhancing the inhomogeneity of the field value until $T \sim T_{\rm c}$, it stops enhancing the inhomogeneity of the energy density of the axion before the amplitude reaches ${\cal O}(1)$. As a result, we find that the direct interaction is not enough to form axion clumps with ${\cal O}(1)$ density contrast, when the self-interaction is negligible and the amplitude of the adiabatic perturbation at $k \sim a_{\rm c} H_{\rm c}$ can be extrapolated from the CMB scales, being ${\cal P}_\zeta \sim 10^{-9}$. The reason why a different conclusion was reached in Ref.~\cite{Sikivie:1982qv} was explained, focusing on the cancellation which should happen for the density perturbation of the axion. 

We also pointed out two possible new scenarios in which axion miniclusters may be potentially formed, even when the PQ symmetry was already broken during inflation. First, when the self-interaction of the axion is not negligible, the parametric resonance can lead to a formation of ${\cal O}(1)$ axion clumps. Second, when the primordial adiabatic perturbation satisfies ${\cal P}_\zeta \geq {\cal O}(10^{-3})$ for $k \sim a_{\rm c} H_{\rm c}$, e.g., as a consequence of a non-trivial evolution during inflation. It has been claimed that when the adiabatic curvature perturbation at $k \sim a_{\rm c} H_{\rm c}$ is ${\cal P}_\zeta \geq {\cal O}(10^{-2})$ for the Gaussian distribution, a significant number of solar mass primordial black holes can be formed (see, e.g., Ref.~\cite{Kalaja:2019uju})\footnote{Since the adiabatic curvature perturbation $\zeta$ is not a local variable, there are some subtle issues in using $\zeta$ as a criterion for the formation of primordial black holes.}. Taking this critical value for granted, our result indicates that if solar mass primordial black holes are one day discovered, we need to take into account the possibility that axion miniclusters are formed also for $f> H_I/2\pi$, which is often called the pre-inflationary scenario\footnote{We should keep in mind that primordial black holes also can be formed without the amplification of the primordial spectrum. For example, as discussed in Ref.~\cite{Garriga:2015fdk}, primordial black holes can be formed also through the collapse of domain walls formed during inflation. Related to this, in the context of the QCD axion, it was pointed out that primordial black holes can be formed from the string-wall network \cite{Ferrer:2018uiu} and QCD axion bubbles \cite{Kitajima:2020kig}.}.

For the first scenario, we need a more careful discussion to understand whether the axion clumps formed around $T \sim T_{\rm c}$ turn to be axion miniclusters because of the two following reasons. First, because of the presence of the self-interaction, the axion clumps can decay through the number changing process likewise the oscillons which are known to be metastable. Second, the resonance peak is not very far away from the Jeans scale. In our forthcoming paper, we will examine these aspects, conducing the non-linear simulation. Meanwhile, in the second scenario, the formation of the axion miniclusters is more likely, since the self-interaction is negligible and the fluctuation of the axion is enhanced to be ${\cal O}(1)$ at larger scales.

In this paper, we have only considered the time evolution around the QCD phase transition. In order to conclude whether axion miniclusters can be indeed formed in these two scenarios, certainly we also need to solve the late time evolution. We also leave this for our future study. Regarding this late time evolution, a number of preceding works can be found for the post inflationary scenario~\cite{Zurek:2006sy, Eggemeier:2019khm, Kavanagh:2020gcy, Xiao:2021nkb}.

In this paper, we simply assumed the radiation domination also during the QCD epoch. The recent progress in lattice QCD simulation has revealed a deviation of the equation of state parameter $w$ and the (squared) sound speed $c_s^2$ from $1/3$ with increased precision (see e.g., Ref.~\cite{Borsanyi:2016ksw}). As discussed e.g., in \cite{Schmid:1998mx}, when we take into account this effect, (axion) dark matter fluctuation tends to grow faster than the radiation dominated epoch. This study is left for elsewhere.

\acknowledgments
Y.~U. would like to thank J.~Garriga for his fruitful comments on the formation of primordial black holes.
\revise{The discussion with P.~Sikivie and W.~Xue was very helpful to understand the difference between our result and theirs.}
N.~K and Y.~U. are supported by Grant-in-Aid for Scientific Research (B) under Contract No.~19H01894 and Fostering Joint International Research (B) under Contract No.~21KK0050. N.~K. is supported by JSPS KAKENHI Grant No. 20H01894, 20H05851, and 21H01078. K.~K. is supported by JSPS KAKENHI Grant No. JP19J22018. Y.~U. is in part supported by the Deutsche Forschungsgemeinschaft (DFG, German Research Foundation) - Project number 315477589 - TRR 211.

\appendix

\section{Importance of the direct interaction between axion and radiation}  \label{App}
Here, we show that the direct interaction between the axion and the QCD matters (radiation), which is described by $V_{\phi \rho} \delta \rho$ in Eq.~(\ref{eq:KGLG}), is crucial both in the superhorizon and subhorizon regimes.  

\subsection{Superhorizon evolution in different gauges}\label{App:diff-gauges}
As discussed in Sec.~\ref{SSec:superhorizon}, the fluctuation of the axion in the uniform radiation density slicing, $\sg$, evolves differently in superhorizon scales from the one in Newtonian gauge, $\delta \phi$. As shown in Eq.~(\ref{Eq:deltaphir}), $\sg$ is decoupled from the adiabatic perturbation in the superhorizon limit, while $\delta \phi$ is not.

\begin{figure}[htbp]
    \centering
    \includegraphics[width=0.49\linewidth]{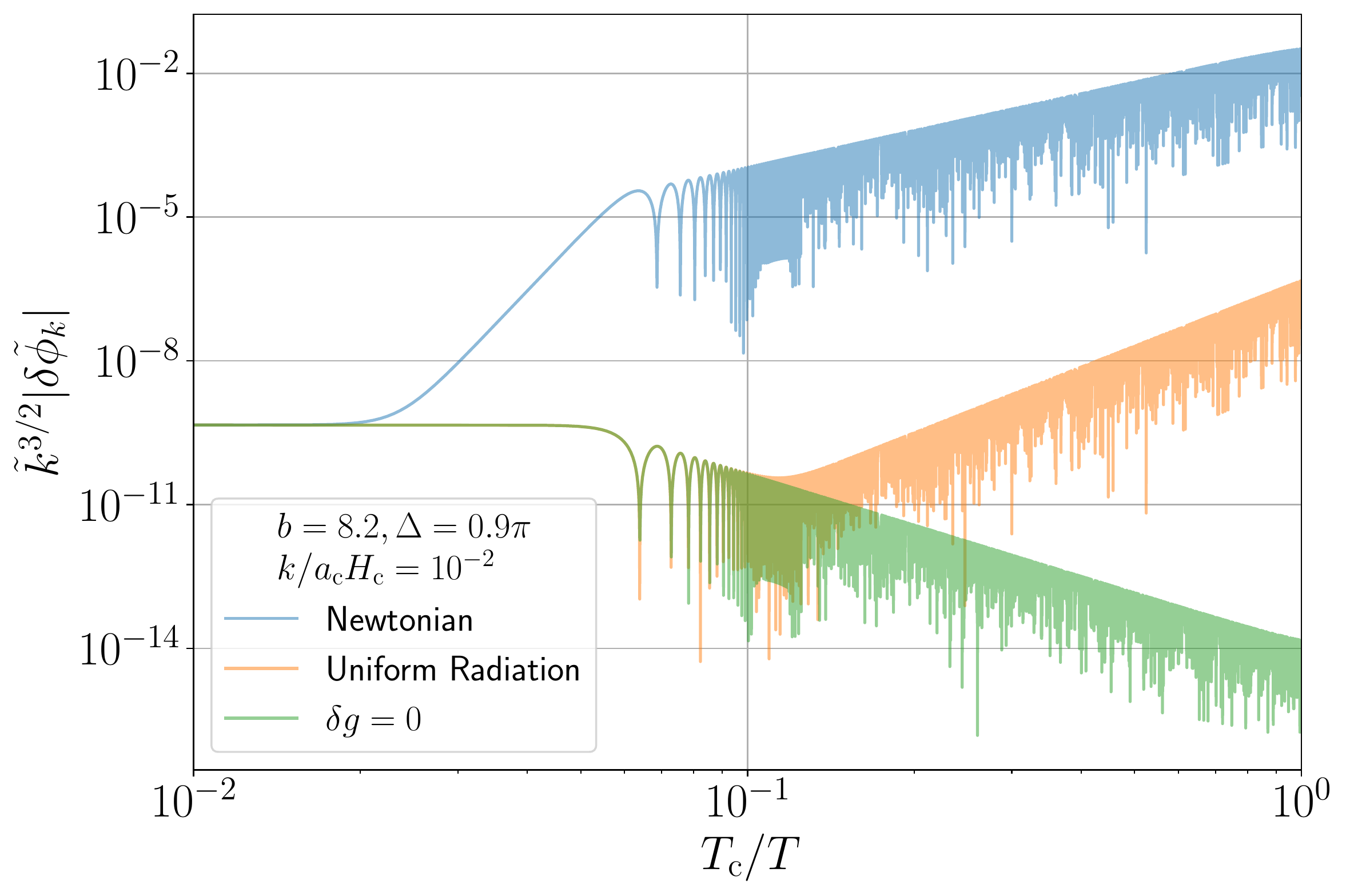}
    \includegraphics[width=0.49\linewidth]{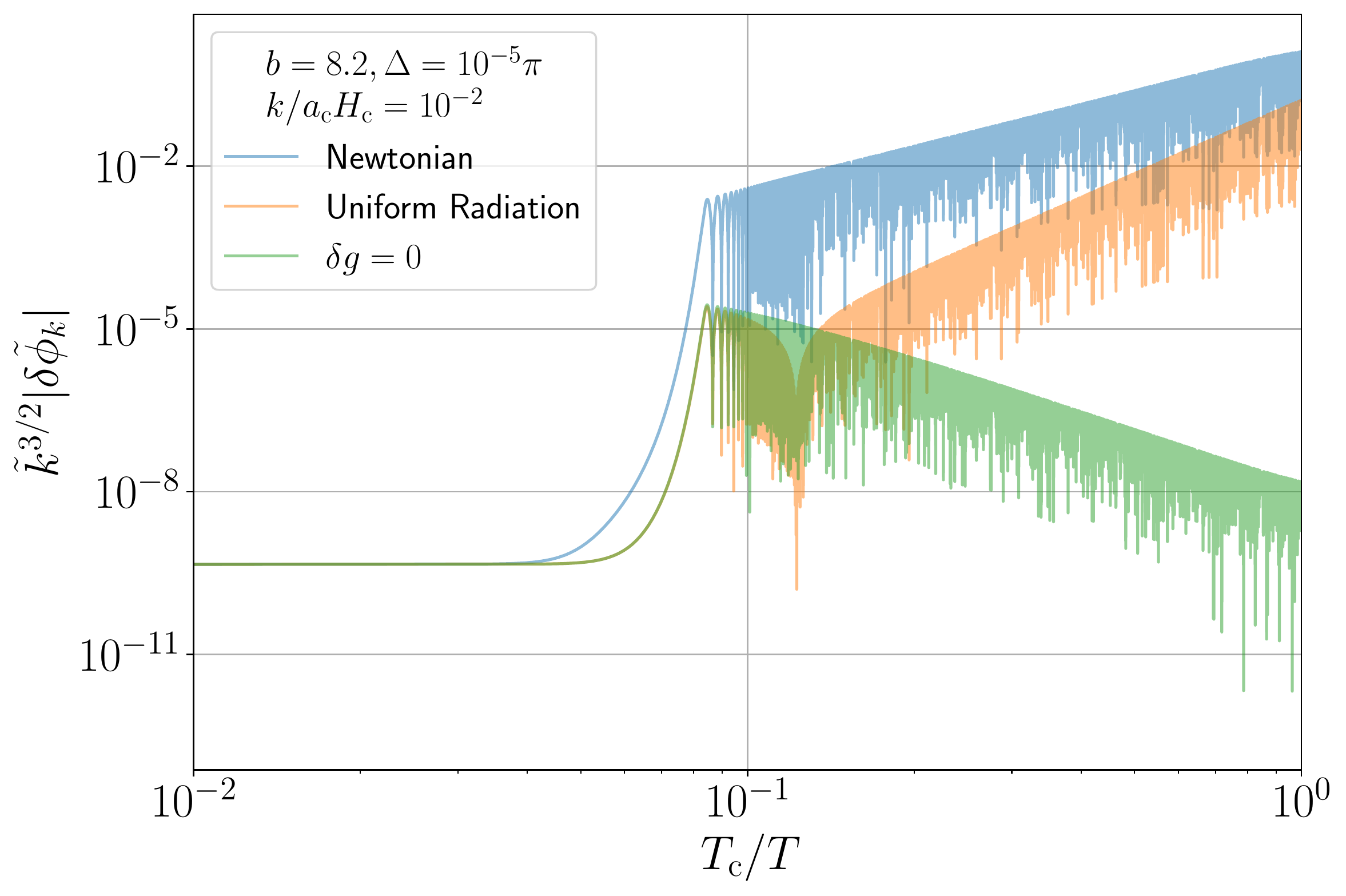}
    \caption{These plots show the time evolution of the axion fluctuation in Newtonian gauge $\delta \phi$, in the uniform radiation density slicing $\sg$, and in the absence of the metric perturbations $\delta \phi_{\delta g=0}$ for $f=10^9\,{\rm GeV}$, $k/(a_{\rm c} H_{\rm c}) = 10^{-2}$, $r_{\rm axi} = 10^{-5}$. The initial background value of the axion is set to $\Delta = 0.9 \pi$ in the left panel and $\Delta = 10^{-5} \pi$ in the right one. These plots show that the evolution of $\sg$ can be precisely traced even if we ignore the adiabatic curvature perturbation at superhorizon scales. In the right panel, these three perturbations are also enhanced by tachyonic instability. 
    \label{fig:dphiSP} }
\end{figure}
Figure \ref{fig:dphiSP} compares the time evolution of $\delta \phi$, $\sg$, and the fluctuation of the axion in the absence of metric perturbations, $\delta \phi_{\delta g=0}$, for a superhorizon mode, $k/(a_{\rm c} H_{\rm c}) = 10^{-2}$. Here, $\sg$ was computed in Newtonian gauge, i.e., by plugging $\delta \phi$ computed by solving Eq.~(\ref{eq:KGLG}) and Eq.~(\ref{Eq:deltarho}) into Eq.~(\ref{Def:sg}). We have chosen the same initial condition for $\sg$ and $\delta \phi_{\delta g=0}$ as $r_{\rm axi} = |\delta \tilde{\phi}_{\rm r}/\zeta^{\rm p}| = |\delta \tilde{\phi}_{\delta g=0}//\zeta^{\rm p}| = 10^{-5}$ and $\partial_t \sg= \partial_t \delta \phi_{\delta g=0}=0$, where we set the primordial amplitude of $\zeta$ as ${\cal P}_{\zeta} = 2.1 \times 10^{-9}$. Using the relation between $\delta \phi$ and $\sg$, we can obtain the corresponding initial condition for $\delta \phi$. The initial background value of the axion is set to $\Delta = 0.9 \pi$ in the left panel and $\Delta = 10^{-5} \pi$ in the right one. The velocity of the background axion was determined by imposing the slow-roll condition. As discussed in  Sec.~\ref{SSec:superhorizon}, $\sg$ and $\delta \phi_{\delta g=0}$ follow the exactly same evolution in the limit $k/(aH) \to 0$. Since $\delta \phi$ is not orthogonal to the adiabatic perturbation $\Phi$, $\delta \phi$ grows rapidly, being enhanced by $\Phi$, once the coefficients of the coupling, $\dot{\phi}$ and $V_\phi$, start to have non-zero values. This happens slightly later in the right panel than in the left panel because of the delayed onset of the oscillation as a consequence of the fine-tuned initial condition.

When we drop the direct interaction, described by $V_{\phi \rho} \delta \rho$ in Eq.~(\ref{eq:KGLG}), the orthogonality between $\sg$ and the adiabatic perturbation $\Phi$ does not properly hold. In fact, without this term, $\sg$ follows a similar evolution to the one for $\delta \phi$, being sourced by $\Phi$, also in superhorizon scales. In the right panel of Fig.~\ref{fig:dphiSP}, since the background initial value is tuned around the hilltop of the potential, all of the three perturbations, $\delta \phi$, $\sg$, and $\delta \phi_{\delta g=0}$ are significantly enhanced by the tachyonic instability.

\subsection{Subhorizon evolution}
\begin{figure}[htbp]
    \centering
    \includegraphics[width=0.49\linewidth]{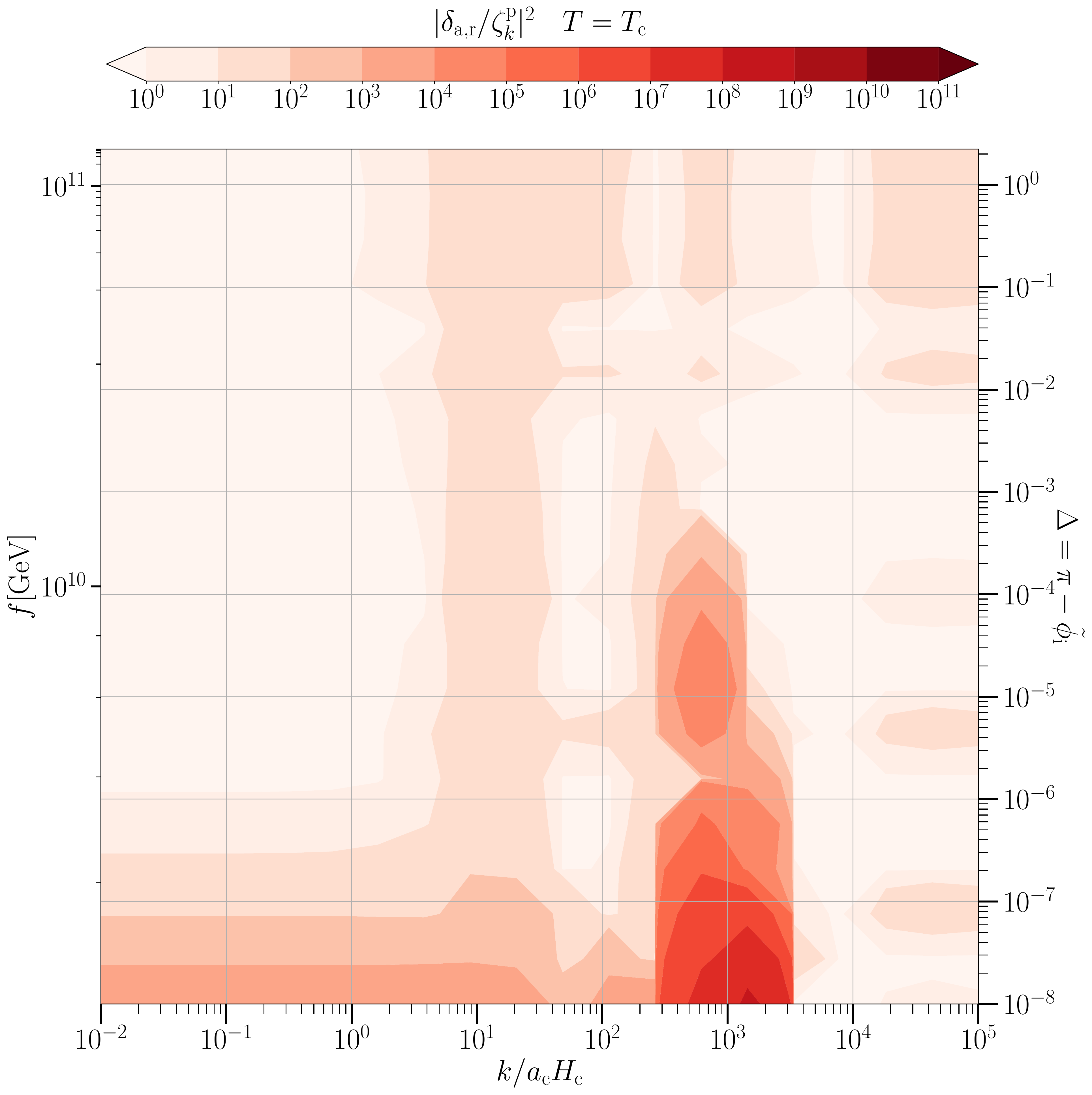}
    \caption{\revise{This figure shows the transfer function of the axion density perturbation which was computed  at $\eta=1$, i.e., at $T=T_{\rm c}$ without including the direct interaction with the radiation for $b=8.2$ and $r_{\rm axi} = 10^{-5}$.}
    Requiring that the axion abundance saturates the total dark matter abundance, we relate the decay constant to the initial misalignment angle as Eq.~(\ref{eq:f-delta}).\label{fig:transfer_noS3}}
\end{figure}
As discussed in Sec.~\ref{SSec:interactions}, the direct interaction between the axion and the radiation, described by $S_3$, becomes the dominant source of the adiabatic perturbation at subhorizon scales. As a reference, in Fig.~\ref{fig:transfer_noS3}, we show the transfer function of the axion density perturbation which was computed without including $S_3$, i.e., in the same setup as Ref.~\cite{Arvanitaki:2019rax}. A comparison between Fig.~\ref{Fg:transferfn} and Fig.~\ref{fig:transfer_noS3} highlights the significance of the direct interaction.

\bibliography{refs.bib}
\end{document}